%% file: arxiv.tex
\documentclass[10pt,twocolumn,amsmath,amssymb,aps,superscriptaddress,pra]{revtex4-2}
\usepackage{amsmath,amssymb}
\usepackage{bm}
\usepackage{graphicx}
\usepackage{caption}
\usepackage{algorithmicx}
\usepackage{algpseudocode}
\usepackage{url}
\usepackage{comment}
\usepackage{tabularx,theorem}
\usepackage{indentfirst}
\usepackage{enumerate}
\usepackage{color}
\usepackage{physics}
\usepackage{mathtools}
\usepackage{qcircuit}
\usepackage{ulem}
\usepackage[colorlinks=true,citecolor=blue,linkcolor=magenta]{hyperref}

\DeclareCaptionType{algorithm}[Algorithm][List of Algorithms]

\newcounter{algorithmfigure}

\DeclareCaptionType{algorithmfigure}[Algorithm][List of Algorithms]
\DeclareCaptionLabelFormat{alglabel}{Algorithm~#2}
\newcommand{\pluseq}{\mathrel{+}=}
\newcommand{\hiddenforreview}[1]{#1}

\newcommand{\Description}[1]{}

\begin{document}

\title{Efficient magic state cultivation with lattice surgery}

\author{Yutaka Hirano}
\affiliation{
    Graduate School of Engineering Science, The University of Osaka, 1-3 Machikaneyama, Toyonaka, Osaka 560-8531, Japan
}

\author{Riki Toshio}
\affiliation{
Quantum Laboratory, Fujitsu Research, Fujitsu Limited,
4-1-1 Kawasaki, Kanagawa 211-8588, Japan
}
\affiliation{
Fujitsu Quantum Computing Joint Research Division,
Center for Quantum Information and Quantum Biology, The University of Osaka, 1-2 Machikaneyama, Toyonaka, Osaka, 565-8531, Japan
}

\author{Tomohiro Itogawa}
\affiliation{
    Graduate School of Engineering Science, The University of Osaka, 1-3 Machikaneyama, Toyonaka, Osaka 560-8531, Japan
}

\author{Keisuke Fujii}
\affiliation{
    Graduate School of Engineering Science, The University of Osaka, 1-3 Machikaneyama, Toyonaka, Osaka 560-8531, Japan
}
\affiliation{
    School of Engineering Science, The University of Osaka, 1-3 Machikaneyama, Toyonaka, Osaka 560-8531, Japan
}
\affiliation{
    Center for Quantum Information and Quantum Biology, The University of Osaka, 1-2 Machikaneyama, Toyonaka 560-0043, Japan
}
\affiliation{
    RIKEN Center for Quantum Computing (RQC), Hirosawa 2-1, Wako, Saitama 351-0198, Japan
}

\date{\today}

\input{0-abstract}
  
\maketitle

\input{1-introduction}
\input{2-preliminary}
\input{3-msc-bottleneck}
\input{4-proposal}
\input{5-performance-evaluation}
\input{6-conclusion}

\section*{Author contributions}
YH designed the protocol and performed the numerical simulations.
RT and TI refined the protocol and suggested the use of lookup tables.
KF supervised the work.
All authors contributed to the discussion of the results and the writing of the manuscript.
Large language models were used to refine the writing of the manuscript and assist in the visualization of the numerical simulation results.

\section*{Acknowledgment}
The authors thank Yosuke Ueno for helpful feedback on the manuscript.
This work is supported by MEXT Quantum Leap Flagship Program (MEXT Q-LEAP) Grant No. JPMXS0120319794, JST COI-NEXT Grant No. JPMJPF2014, JST Moonshot R\&D Grant No. JPMJMS2061, and JST CREST JPMJCR24I3.

\clearpage
\bibliographystyle{unsrt}
\bibliography{refs}

\end{document}

%% file: 0-abstract.tex
\begin{abstract}
Magic state distillation plays a crucial role in fault-tolerant quantum computation and represents a major bottleneck.
In contrast to traditional logical-level distillation, physical-level distillation offers significant overhead reduction by enabling direct implementation with physical gates.
Magic state cultivation is a state-of-the-art physical-level distillation protocol that is compatible with the square-grid connectivity and yields high-fidelity magic states.
However, it relies on the complex grafted code, which incurs substantial spacetime overhead and complicates practical implementation.
In this work, we propose an efficient cultivation-based protocol compatible with the square-grid connectivity.
We reduce the spatial overhead by avoiding the grafted code and further reduce the average spacetime overhead by utilizing code expansion and enabling early rejection.
Numerical simulations show that, with a color code distance of 3 and a physical error probability of $10^{-3}$, our protocol achieves a logical error probability for the resulting magic state comparable to that of magic state cultivation ($\approx 3 \times 10^{-6}$), while requiring about half the spacetime overhead.
Our work provides an efficient and simple distillation protocol suitable for megaquop use cases and early fault-tolerant devices.
\end{abstract}

%% file: 1-introduction.tex
\section{Introduction}
\label{sec:introduction}
Quantum computers promise to solve problems of practical importance that are intractable for classical computers, such as simulating quantum systems~\cite{lloyd1996universal} and integer factoring~\cite{Shor1994Factoring}.
Existing quantum computers are highly susceptible to noise, which prevents them from executing complex algorithms such as Shor's factoring algorithm.
To execute such complex quantum algorithms, we need fault-tolerant quantum computers (FTQC)~\cite{shor1996fault}, which rely on quantum error correction (QEC)~\cite{shor1995scheme, Steane1996ErrorCorrecting}.

In FTQC, each \textit{logical gate} must be implemented in a fault-tolerant manner.
While fault-tolerant implementations of Clifford gates are relatively inexpensive, Clifford gates alone are not sufficient for universal quantum computation~\cite{gottesman1998heisenberg}.
Implementing a non-Clifford gate such as the $T$ gate requires a special quantum state, a \textit{magic state}, which is obtained through a process known as \textit{magic state distillation}~\cite{Bravyi2005}.
Magic state distillation is resource-intensive and remains a major bottleneck in fault-tolerant quantum computation.
Consequently, many distillation protocols have been proposed~\cite{Bravyi2005, knill1998resilient, bravyi2012magic, meier2012magic, Goto2016Minimizing, Campbell2017Unified, Gidney2019EfficientMagicState, Litinski2019GameOfSurfaceCodes, Litinski2019magicstate, chamberland2020very, itogawa2025efficient, Hirano2024Leveraging, Gidney2024MagicStateCultivation, vaknin2025magic, chen2025efficient, sahay2025fold, claes2025cultivating}.

Bravyi~\textit{et al.}~\cite{Bravyi2005} proposed the 15-to-1 distillation protocol, which exploits $T$'s transversality on the $[[15, 1, 3]]$ code.
Earlier, predating the term ``magic state distillation'', Knill~\textit{et al.}~\cite{knill1998resilient} proposed a protocol based on the Hadamard test of the $H$ gate, relying on $H$'s transversality on the $[[7, 1, 3]]$ code.
These studies formulated magic state distillation as a quantum algorithm built from ideal Clifford gates with noisy inputs, and this formulation was followed by subsequent studies~\cite{bravyi2012magic, meier2012magic, Campbell2017Unified, Gidney2019EfficientMagicState, Litinski2019GameOfSurfaceCodes}.
Running such algorithms on FTQC requires protecting each gate against noise using QEC, incurring a significant cost.

Goto~\cite{Goto2016Minimizing} proposed a technique that we call \textit{physical-level distillation}.
In contrast to the above formulation, the distillation circuit operates with noisy Clifford gates.
Specifically, the circuit detects any single gate error, thereby achieving a logical error probability of $\mathcal{O}(p_\mathrm{phys}^2)$, where $p_\mathrm{phys}$ is the physical error probability.
To this end, the distillation protocol consists of a Knill-style Hadamard test with a flag qubit.
Its physical-gate implementation significantly reduces the distillation overhead.

Goto's protocol requires all-to-all connectivity.
Chamberland~\textit{et al.}~\cite{chamberland2020very} and Itogawa~\textit{et al.}~\cite{itogawa2025efficient} proposed physical-level distillation protocols that work with planar connectivity.
The latter, known as zero-level distillation, operates on the simpler square-grid connectivity and yields a magic state encoded in a surface code~\cite{Kitaev2003, bravyi2007measurement}, achieving a logical error probability of $10^{-4}$ for a realistic $p_\mathrm{phys} = 10^{-3}$ setting.
Gidney~\textit{et al.}~\cite{Gidney2024MagicStateCultivation} refined these works and proposed \textit{magic state cultivation} (MSC), which works with the square-grid connectivity.
This protocol extensively utilizes postselection to suppress errors, thereby producing high-fidelity magic states.
As a result, it yields a magic state with a logical error probability of $2 \times 10^{-9}$ at $p_\mathrm{phys} = 10^{-3}$.
MSC prepares a magic state encoded in a 2D color code~\cite{bombin2006topological}, which is then transformed into the grafted code, another QEC code compatible with matching QEC decoders.
However, due to its high-weight stabilizers, the grafted code is less space-efficient than the rotated surface code~\cite{bombin2007optimal} with a comparable logical error probability, incurring substantial spacetime overhead.

Although these physical-level distillation protocols significantly reduce the distillation overhead, further advances are desirable, particularly when many magic states are consumed simultaneously during the parallel execution of multiple $T$ gates~\cite{Beverland2022Assessing, hirano2025locality}.
While this has motivated the development of various MSC-based protocols~\cite{vaknin2025magic, chen2025efficient, sahay2025fold, claes2025cultivating}, they relax the connectivity restriction that MSC respects.
This restriction is particularly important for superconducting devices.
Additionally, although platforms such as neutral atoms allow flexible connectivity, qubit shuttling incurs temporal and idling costs; thus, maintaining local operations on a fixed lattice can offer temporal advantages.
Therefore, there is a need for a more efficient magic state distillation protocol compatible with the square-grid connectivity.

In this study, we propose a novel MSC-based protocol that significantly reduces the spacetime overhead of distillation without relaxing the hardware connectivity requirements.
We avoid using the grafted code by employing lattice surgery~\cite{Horsman2012, poulsen2017fault}, as in the original zero-level distillation~\cite{itogawa2025efficient}, to transfer the magic state from the color code to the rotated surface code.
Lattice surgery and the rotated surface code allow us to employ two techniques to further reduce the distillation overhead:
(1) utilizing surface code expansion~\cite{li2015magic} to minimize the qubit count in the early stages of distillation, and (2) predicting the logical error probability using a simple lookup table, allowing early rejection to reduce the average spacetime overhead.

In addition to efficiency, our protocol provides simplicity.
The grafted code is complex, and avoiding its implementation is appealing to practitioners.
The rotated surface code and lattice surgery are simpler, already supported by existing compilers~\cite{Watkins2023High} and instruction sets, and impose little additional burden on control hardware.
Thus, our protocol can leverage existing software and hardware infrastructures, making it more practical for near-term implementations.

We conducted numerical simulations using Stim~\cite{gidney2021stim} to evaluate the performance of our protocol.
With a color code distance of 3, our protocol achieves a logical error probability for the resulting magic state comparable to that of MSC, while requiring about half the spacetime overhead.
Specifically, at $p_\mathrm{phys} = 10^{-3}$, our protocol achieves a logical error probability of $1.6 \times 10^{-6}$ for $S\ket{+}$ distillation.
This is comparable to the $1.2 \times 10^{-6}$ achieved by MSC with color code distance 3, and corresponds to a logical error probability of $3 \times 10^{-6}$ for $T\ket{+}$ distillation.
For logical error probabilities in the range $[10^{-6}, 10^{-5}]$, our protocol requires less than half the spacetime overhead needed for MSC.
These results demonstrate that our approach offers a practical path toward more efficient magic state distillation in fault-tolerant quantum computation, especially for megaquop use cases~\cite{preskill2025beyond} (computation involving $\sim 10^{6}$ logical gate operations) and early fault-tolerant devices~\cite{akahoshi2024partially,toshio2025practical, akahoshi2024compilation} (systems with $10^3$--$10^5$ physical qubits).

Our contributions are summarized as follows:
\begin{enumerate}
 \item We demonstrate that, although the original MSC work found lattice-surgery-based constructions unsuitable, lattice surgery can in fact be used within MSC-style distillation to realize a simpler and more efficient distillation protocol.
 \item We define a concrete protocol, magic state cultivation with lattice surgery (MSC-LS), an efficient and simple distillation protocol using lattice surgery and surface code expansion. This protocol is compatible with the square-grid connectivity and avoids the use of the grafted code.
 \item We introduce an early-rejection mechanism based on a lookup table of intermediate error syndrome patterns. This mechanism reduces the expected spacetime overhead by aborting attempts that are unlikely to pass the final postselection.
 \item We quantitatively evaluate MSC-LS and show that it achieves logical error probabilities comparable to those of MSC while reducing the expected spacetime overhead by roughly 50\%.
\end{enumerate}

The rest of this paper is structured as follows.
In \autoref{sec:preliminary}, we provide basic definitions and prior work essential for understanding our proposal.
Specifically, we define the detector error model and provide a summary of the complementary gap and magic state cultivation.
In \autoref{sec:msc-bottleneck}, we discuss bottlenecks in magic state cultivation that motivate our proposal.
In \autoref{sec:msc-ls}, we detail our proposal: magic state cultivation with lattice surgery.
In \autoref{sec:performance-evaluation}, we present a performance evaluation of the protocol, including the numerical simulation settings and results.
Finally, \autoref{sec:conclusion} summarizes our findings.

%% file: 2-preliminary.tex
\section{Preliminary}
\label{sec:preliminary}

\subsection{Detector error model}
\label{subsec:preliminary-detector-error-model}
Quantum error correction involves two procedures: \textit{syndrome extraction} (or \textit{syndrome measurement}) and \textit{decoding}.
Syndrome extraction runs quantum circuits (\textit{syndrome extraction circuits}) to gather error information (\textit{error syndrome}).
Decoding is a classical computation process that identifies the locations of errors from the error syndrome.
In FTQC, syndrome extraction circuits are also noisy, and the decoding process must diagnose errors in both the syndrome extraction circuits and the data qubits.
The concepts of detectors and detector error models~\cite{McEwen2023RelaxingHardware, derks2024designing} are useful in such circumstances, and we therefore introduce them here.

\begin{figure}[t!]
\centering
\[
\Qcircuit @C=.4em @R=.8em {
 \lstick{a: \ket{0}} & \targ     & \targ     & \meter & \push{\ket{0}} & \targ     & \targ     & \meter \\
 \lstick{0: \ket{+}} & \ctrl{-1} & \qw       & \qw    & \qw            & \ctrl{-1} & \qw       & \qw \\
 \lstick{1: \ket{+}} & \qw       & \ctrl{-2} & \qw    & \qw            & \qw       & \ctrl{-2} & \qw    
}
\]
\caption{
An example of a detector.
}
\label{fig:preliminary-detector}
\Description{Two-round syndrome extraction circuit for measuring a two-qubit Z stabilizer on data qubits 0 and 1 with an ancilla qubit a.
Although each ancilla measurement outcome can vary, the parity of the two outcomes is deterministic when no error occurs.
The figure illustrates that this deterministic parity is called a detector.}
\end{figure}

A \textit{detector} is defined as the parity of measurement outcomes that is deterministic in the absence of errors.
For example, \autoref{fig:preliminary-detector} illustrates two rounds of syndrome extraction.
Each syndrome extraction circuit performs an indirect measurement of $Z_0Z_1$.
Although the individual measurement outcomes are non-deterministic, the parity of these outcomes is always 0 in the absence of errors and thus forms a detector.
We say that a Pauli error is detected by a detector if the error changes the parity.
For example, in \autoref{fig:preliminary-detector}, an $X$ error on qubit 0 occurring between the two rounds is detected by the detector.

A \textit{Tanner graph}~\cite{tanner2003recursive,loeliger2004introduction} is a useful data structure for capturing the error structure of a syndrome extraction circuit.
A Tanner graph is a bipartite graph $(U, V, E)$ where $U$ and $V$ are disjoint sets of vertices, and $E$ is the set of edges between $U$ and $V$.
In our context, $U$ represents errors, $V$ represents detectors, and an edge exists between an error and a detector if and only if the detector detects the error.
We call a probability assignment to all errors a \textit{noise model}.
A \textit{detector error model} is defined as a Tanner graph together with a noise model.
A QEC decoder diagnoses the locations of errors given a detector error model and an error syndrome.

Matching decoders~\cite{brown2023conservation} require that each error with nonzero probability be detected by at most two detectors in the detector error model.
Such a detector error model is therefore said to be compatible with matching decoders.
For example, the detector error model for the typical syndrome extraction circuit of a rotated surface code is compatible with matching decoders, whereas the detector error model for the typical syndrome extraction circuit of a 2D color code is not.

\subsection{Complementary gap}
\label{subsec:preliminary-complementary-gap}
\begin{figure}[t!]
\centering
\includegraphics[width=7.5cm]{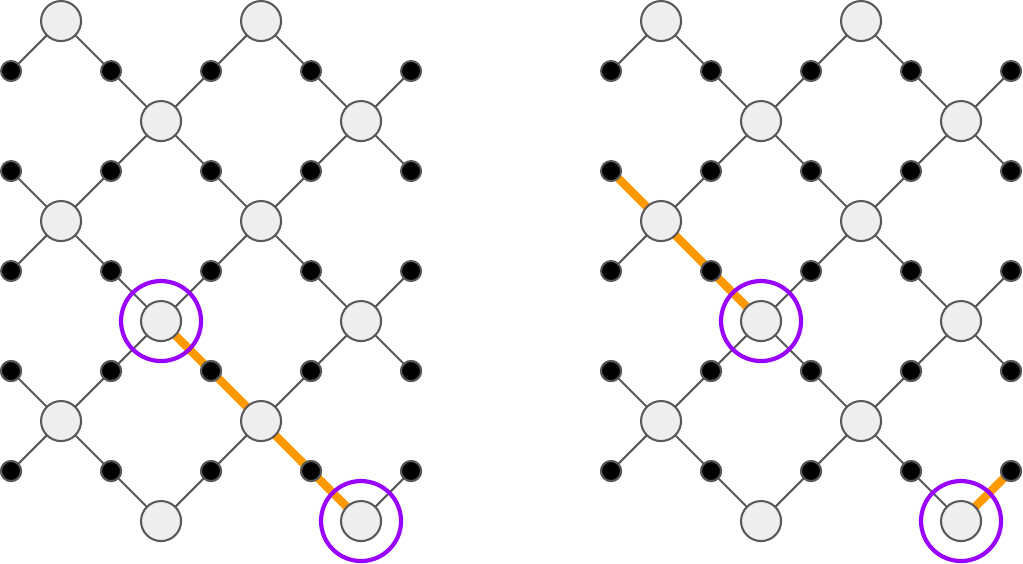}
\caption{
A Tanner graph of the distance-5 rotated surface code (for $Z$ errors only) with an error syndrome.
Gray, black, and purple circles indicate detectors, error locations, and detectors that detect $Z$ errors, respectively.
The left and right panels show matching results that are complementary to each other, with matching paths colored orange.
}
\label{fig:preliminary-complementary-gap}
\Description{Fully described in the text.}
\end{figure}

The minimum-weight perfect matching (MWPM) decoder~\cite{fowler2012towards} maps the decoding problem to a graph matching problem and then solves it using Edmonds' algorithm~\cite{edmonds1965paths, edmonds1965maximum}.
Specifically, it finds a correction with minimum log-likelihood weight, where the log-likelihood weight of a correction is the sum of the prior log-likelihood ratios of the individual error mechanisms included in that correction.
The \textit{complementary gap}~\cite{gidney2025yoked} quantifies the decoder's confidence in its decoding result given an error syndrome, and it can be computed using the MWPM decoder.
In this subsection, for simplicity, we consider only $Z$ errors.
We also assume that syndrome extraction circuits are noise-free, and each syndrome measurement forms a detector.
\autoref{fig:preliminary-complementary-gap} shows the Tanner graphs of the distance-5 rotated surface code under these assumptions.
Each $X$ stabilizer check corresponds to a detector (gray circle), and each data qubit corresponds to an error location (black circle).

The MWPM decoder solves the matching problem to identify error locations.
\autoref{fig:preliminary-complementary-gap} shows two matching solutions, with matching paths colored orange.
Both are valid matching results, and they are complementary to each other in the sense that the union of the matching paths forms a logical $Z$ chain.
If the two solutions are equally likely, the decoder’s confidence in the decoding result is low.

\begin{figure}[t!]
\centering
\includegraphics[width=7.5cm]{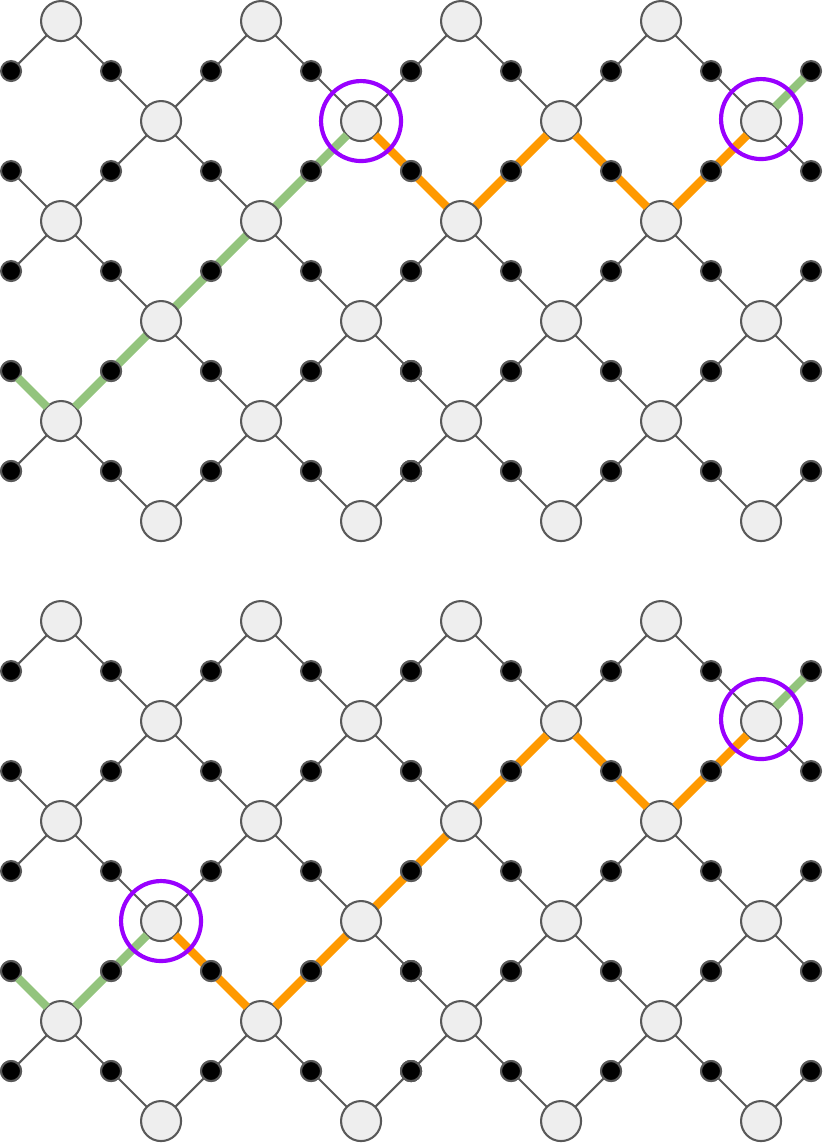}
\caption{
Examples of error syndrome patterns that yield low and high decoding confidence.
The Tanner graphs are shown for a rotated surface code with only $Z$ errors.
Gray, black, and purple circles indicate detectors, error locations, and detectors that detect $Z$ errors, respectively.
The orange and green paths represent candidate correction paths in different logical classes.
(Top) The two candidate paths have similar lengths, resulting in similar weights and a low complementary gap.
(Bottom) The orange path is much longer than the green path, resulting in a high complementary gap.
}
\label{fig:preliminary-complementary-gap-confidence-examples}
\Description{Fully described in the text.}
\end{figure}

The complementary gap quantifies how strongly the decoder favors one logical class over its complementary class.
It is defined as $|w(1) - w(2)|$, where $w(1)$ is the log-likelihood weight of the best correction conditioned on one logical class and $w(2)$ is the log-likelihood weight of the best correction conditioned on the complementary logical class.
To compute this value, the MWPM decoder is run twice with modified detector error models~\cite{hutter2014efficient}, once for a logical class and once for its complementary class.
The complementary gap can be used for soft-output decoding of concatenated codes~\cite{gidney2025yoked} and for postselection in resource preparation to lower the logical error probability~\cite{Gidney2024MagicStateCultivation, vaknin2025magic, chen2025efficient, sahay2025fold, claes2025cultivating, sunami2025entanglement}.
In this work, we use the complementary gap as the final postselection criterion.
Specifically, a distillation attempt is rejected when the complementary gap falls below the chosen threshold.
\autoref{fig:preliminary-complementary-gap-confidence-examples} shows two syndrome examples that yield low and high complementary gaps, respectively.
In the top panel, the orange and green paths contain similar numbers of error locations, leading to similar weights and a low complementary gap; such a low-confidence case would be rejected by postselection.
By contrast, in the bottom panel, the orange path contains many more error locations than the green path, leading to a high complementary gap; such a high-confidence case would pass postselection.

\subsection{Magic state cultivation}
\label{subsec:preliminary-magic-state-cultivation}

\begin{figure}[t!]
\centering
\includegraphics[width=8cm]{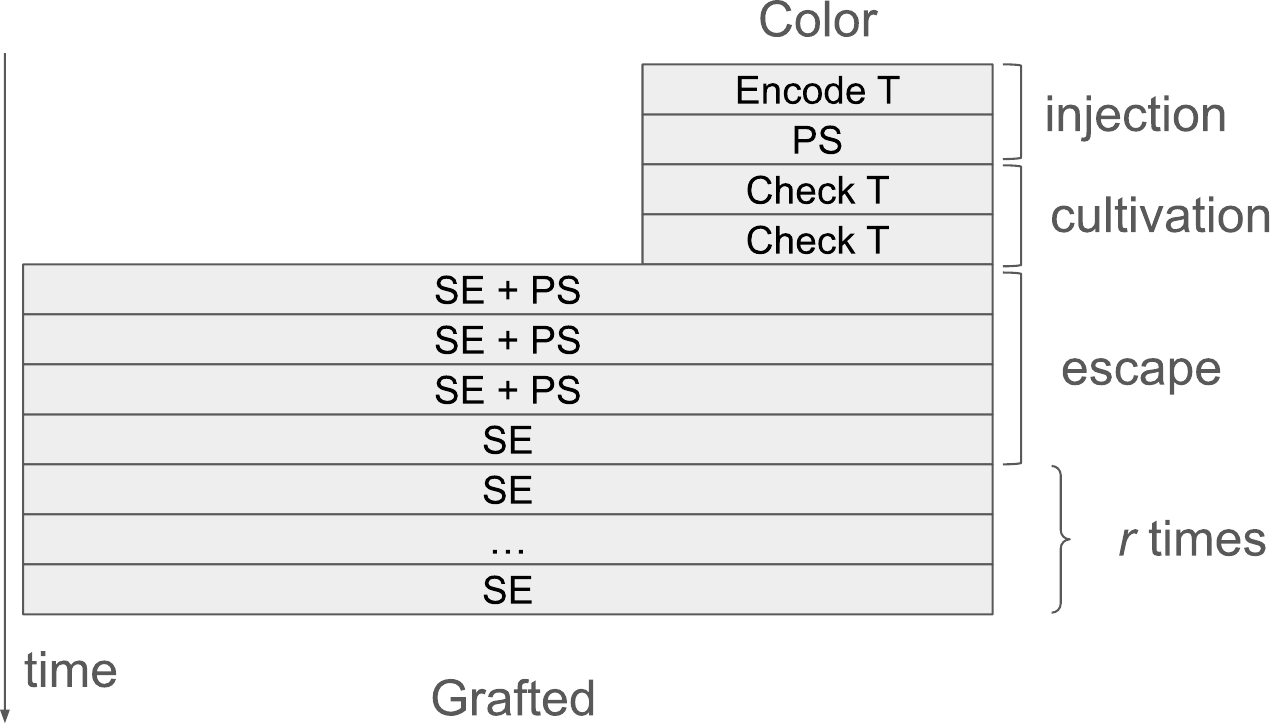}
\caption{
Magic state cultivation~\cite{Gidney2024MagicStateCultivation} for $d_{\mathrm{color}} = 3$.
``Encode T'' denotes the non-fault-tolerant encoding of a magic state into the distance-3 2D color code.
Two consecutive ``Check T'' represent the application of two Hadamard tests.
``PS'' denotes postselection, and ``SE'' denotes syndrome extraction.
}
\label{fig:preliminary-msc-protocol}
\Description{Timeline-style schematic of magic state cultivation for d_color equal to 3.
A vertical arrow indicates the forward progression of time.
The diagram is divided into three labeled stages: injection at the top, cultivation in the middle, and escape at the bottom.
The injection stage consists of an Encode T block followed by a postselection block.
The cultivation stage consists of two Check T blocks.
The escape stage consists of three blocks, each representing syndrome extraction and postselection.
After the escape stage, r more syndrome extraction blocks follow.}
\end{figure}

Magic state cultivation is an efficient magic state distillation protocol.
It consists of three stages: injection, cultivation, and escape.
In the injection stage, a magic state is injected non-fault-tolerantly into the distance-3 2D color code.
In the cultivation stage, the fault distance of the magic state is increased to $d_{\mathrm{color}}$.
Here, we define the \textit{fault distance} of a circuit or a state as the minimum number of errors required to cause a logical error without being detected by any detector.
In the escape stage, the color code is transformed into the grafted code, another QEC code compatible with matching decoders.
Finally, we compute the complementary gap and perform postselection based on its value.
We perform $r$ rounds of syndrome extraction while waiting for the QEC decoder to compute the complementary gap.
\autoref{fig:preliminary-msc-protocol} depicts the protocol for $d_{\mathrm{color}} = 3$.

\subsubsection{Injection stage}
\label{subsec:preliminary-magic-state-cultivation-injection}
In the injection stage, similar to the preceding physical-level distillation protocols~\cite{Goto2016Minimizing, itogawa2025efficient}, a magic state is injected non-fault-tolerantly into the distance-3 2D color code, also known as the Steane seven-qubit code~\cite{Steane1996ErrorCorrecting}.
The original magic state cultivation proposal provides three injection methods: unitary injection, teleport injection, and Bell injection.
In this work, we use unitary injection, which is an injection circuit composed of unitary gates, throughout because it has the best performance of the three.
After the magic state is injected, we perform one round of syndrome extraction followed by postselection; if the error syndrome is nontrivial, we reject the distillation attempt.
At the end of this stage, a magic state with fault distance 1 is encoded in the distance-3 2D color code.

\subsubsection{Cultivation stage}
\label{subsec:preliminary-magic-state-cultivation-cultivation}
In the cultivation stage, the fault distance of the magic state is increased.
Conceptually, this is done by performing Hadamard tests, as in the preceding proposals using physical-level distillation~\cite{Goto2016Minimizing, chamberland2020very, itogawa2025efficient}.
This relies on two facts: (1) the magic state $\ket{T} = T\ket{+}$ is the +1 eigenstate of $H_{XY} = (X + Y) / \sqrt{2}$, and (2) 2D color codes admit transversal $H_{XY}$.
Magic state cultivation employs an efficient implementation of a pair of Hadamard tests (corresponding to the two ``Check T'' boxes in \autoref{fig:preliminary-msc-protocol}) that is well-suited to the square-grid connectivity.

When $d_\mathrm{color} = 3$, the cultivation stage ends here.
With a larger $d_\mathrm{color}$, the 2D color code is expanded and the code distance increases by 2.
After performing syndrome extraction and postselection for code stabilization, another pair of Hadamard tests is performed to increase the fault distance.
This procedure is repeated until the fault distance reaches $d_\mathrm{color}$.

At the end of this stage, a magic state with fault distance $d_\mathrm{color}$ is encoded in the 2D color code of distance $d_\mathrm{color}$.
Similar to the injection stage, in this stage, any nontrivial measurement outcome leads to rejection of the distillation attempt.

\subsubsection{Escape stage}
\label{subsec:preliminary-magic-state-cultivation-escape}
In the escape stage, the 2D color code of distance $d_\mathrm{color}$ is transformed into the grafted code of distance $d_\mathrm{grafted}$, another QEC code compatible with matching decoders.
The transformation involves three steps:
(1) creating a combination of a surface code and a color code,
(2) stabilizing the code by measuring its stabilizers several times, and
(3) making the color code region matchable by dropping some stabilizers and decomposing others into two-body stabilizers.
During the stabilization step, we manually perform postselection for some detectors on the color code region and remove them from the detector error model used by the QEC decoder to keep the detector error model compatible with matching decoders.

Because the grafted code is compatible with matching decoders, we can compute the complementary gap and perform postselection based on it.
Note that the fault distance of the resulting magic state depends on the complementary gap threshold.
The grafted code of distance $d$ is similar to the rotated surface code of distance $d$ but has a higher logical error probability.
For example, Gidney~\textit{et al.}~\cite{Gidney2024MagicStateCultivation} report that the logical error probability of the grafted code of distance 15 is comparable to that of the rotated surface code of distance 11.

%% file: 3-msc-bottleneck.tex
\section{Bottlenecks in magic state cultivation} \label{sec:msc-bottleneck}

\begin{figure}[t!]
\centering
\includegraphics[width=8.5cm]{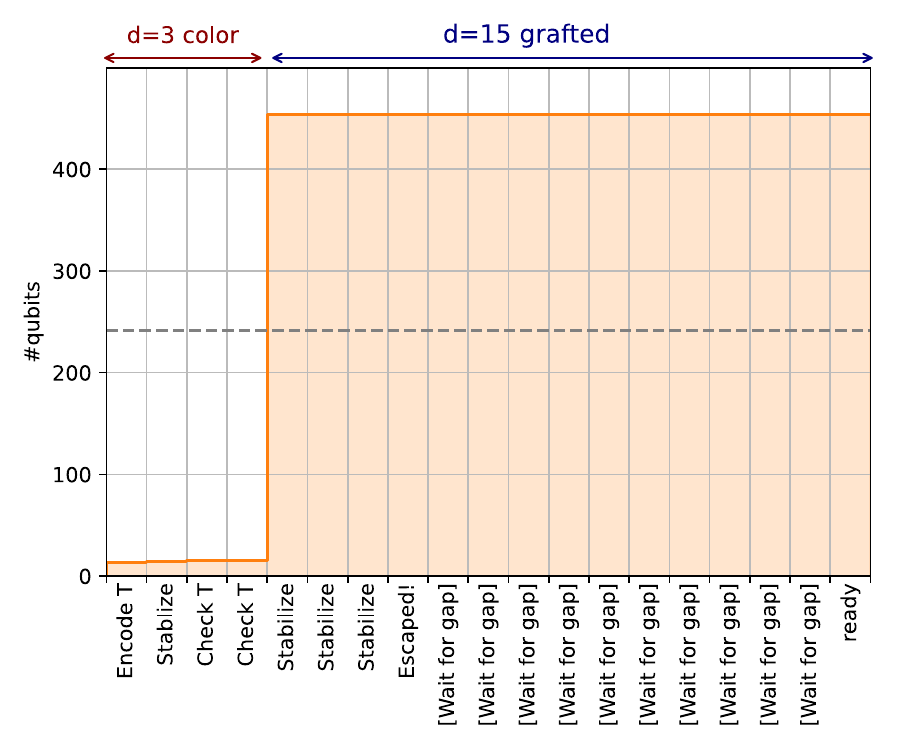}
\caption{
Qubit usage during magic state cultivation with $d_\mathrm{color} = 3$, $d_\mathrm{grafted} = 15$, and $r = 10$.
Each column represents a round.
See \autoref{subsec:performance-evaluation-simulation-settings} for the definition of ``round''.
The dashed line represents the number of physical qubits required for the distance-11 rotated surface code.
}
\label{fig:msc-bottleneck-qubit-count}
\Description{Step plot of qubit usage across the rounds of magic state cultivation.
The horizontal axis lists the stages from Encode T through Stabilize, Check T, Escaped, repeated wait-for-gap rounds, and ready.
The vertical axis shows the number of qubits.
The plot begins at a low qubit count during the color-code stage, then rises sharply at the transition to the grafted code and remains flat at a much higher level, 454 physical qubits, through the remaining rounds.
A horizontal dashed line marks the qubit count of a distance-11 rotated surface code, 241 physical qubits, and lies well below the long high plateau of the grafted-code stage.}
\end{figure}

In this section, we discuss bottlenecks in magic state cultivation.
\autoref{fig:msc-bottleneck-qubit-count} shows the qubit usage during distillation with parameters described later in \autoref{sec:performance-evaluation}.
Specifically, we use parameters $d_\mathrm{color} = 3$, $d_\mathrm{grafted} = 15$, and $r = 10$.
As shown in the figure, the spatial cost of the distance-15 grafted code, which is 454 qubits, is dominant.
Moreover, we need to wait $r = 10$ rounds for the QEC decoder to compute the complementary gap, which increases the impact of the final code's spatial cost on the total spacetime overhead.
Gidney~\textit{et al.}~\cite{Gidney2024MagicStateCultivation} report that the logical error probability of the grafted code of distance 15 is comparable to that of the rotated surface code of distance 11.
The distance-11 rotated surface code requires 241 qubits (the dashed line in \autoref{fig:msc-bottleneck-qubit-count}); hence, if we were able to replace the distance-15 grafted code with the distance-11 rotated surface code---as our proposal later enables---the impact would be substantial.

As discussed in \autoref{sec:introduction}, the grafted code is complex, making its implementation unattractive to practitioners.
A key issue is that this complexity persists beyond the distillation process: the magic state encoded in the grafted code remains until it is consumed by gate teleportation.
While the distillation process is isolated from the rest of the system, once distillation completes the grafted code patch storing the magic state participates in the system's normal error-correction cycle.
Consequently, this additional architectural complexity is more problematic than complexity confined to the distillation process.

More concretely, supporting both the grafted code and the rotated surface code affects multiple layers of the control stack.
First, the syndrome extraction schedules for the two codes must be synchronized, as analyzed in Ref.~\cite{maurya2025synchronization}.
Second, real-time decoding must support both QEC codes.
For example, architectures based on extensive lattice surgery~\cite{Litinski2019GameOfSurfaceCodes} are expected to rely on parallel-window decoding in practical systems.
Supporting both codes makes implementing parallel-window decoding more challenging.
Moreover, in compilation schemes that allow distillation factories to be placed at arbitrary locations~\cite{hirano2025locality}, the system must support the grafted code at arbitrary locations as well, further increasing the implementation burden.
Avoiding such additional architectural complexity is therefore desirable.

In some architectures where the rotated surface code is used for computation, supporting the rotated surface code as the final code would provide additional benefits beyond those discussed above.
For example, Zhou~\textit{et al.}~\cite{zhou2025resource} use magic state cultivation in their transversal architecture for reconfigurable atom arrays.
To enable transversal CX for gate teleportation, they transform the grafted code into the rotated surface code by expanding the grafted code and cutting away the color code region.
This additional conversion incurs extra spatial cost---an overhead that would be unnecessary if the final code were the rotated surface code instead of the grafted code.

\begin{figure}[t!]
\centering
\includegraphics[width=8.5cm]{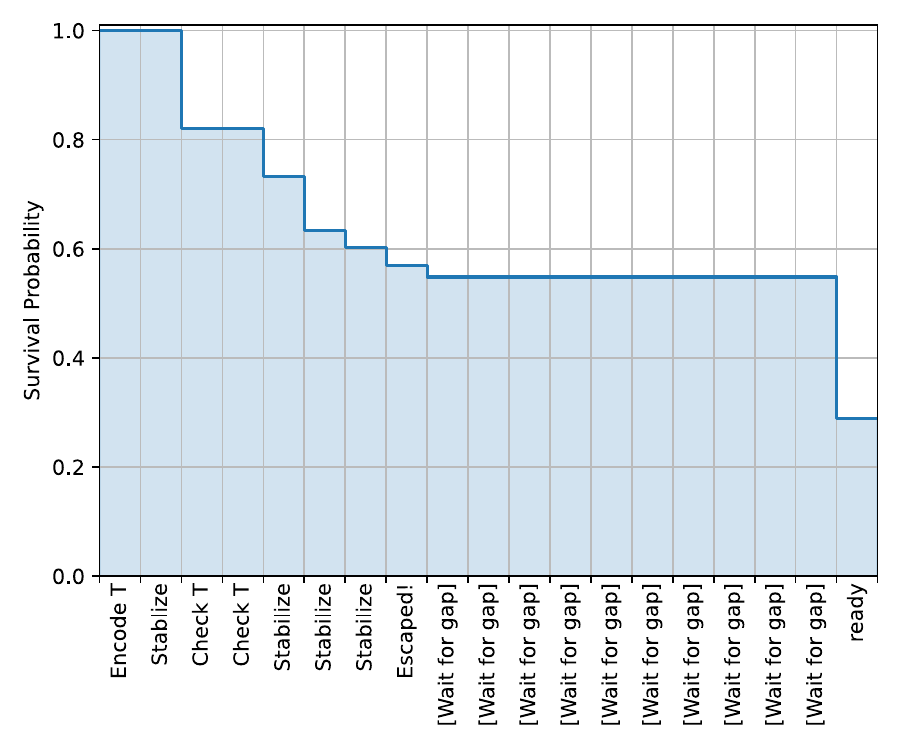}
\caption{
Survival probability during magic state cultivation with $d_\mathrm{color} = 3$, $d_\mathrm{grafted} = 15$, and $r = 10$.
The logical error probability of the resulting $S\ket{+}$ state is $1.6 \times 10^{-6}$.
Each column represents a round.
}
\label{fig:msc-bottleneck-survival-probability}
\Description{Step plot of survival probability across the rounds of magic state cultivation.
The horizontal axis lists the stages from Encode T through Stabilize, Check T, Escaped, repeated wait-for-gap rounds, and ready.
The vertical axis shows survival probability from 0 to 1.
The plot starts at 1.0, decreases in several steps during the early rounds, stays nearly flat at about 0.55 through the repeated wait-for-gap rounds, and then drops again at the final ready stage to about 0.29.
The blue shaded region marks the surviving fraction after each round.}
\end{figure}

Another important metric is the cumulative success probability across rounds, as illustrated in \autoref{fig:msc-bottleneck-survival-probability}.
The blue region in the figure represents the proportion of attempts surviving up to each round of the distillation process.
For example, the end-to-end success probability of the distillation process in the figure is 29\%.
While the proportion of rejected attempts is important, the timing of rejection also matters.
In the figure, 26\% of attempts are rejected by the final postselection.
Rejecting those attempts in earlier stages would enable retrying distillation earlier and thus reduce the average spacetime cost of distillation.

Following the proposal of magic state cultivation, several related protocols have been proposed~\cite{vaknin2025magic, chen2025efficient, sahay2025fold, claes2025cultivating}.
These protocols rely on non-local connectivity to perform the cultivation stage on QEC codes other than 2D color codes.
For example, Refs.~\cite{sahay2025fold, claes2025cultivating} use unrotated surface codes, and Ref.~\cite{chen2025efficient} uses SRP codes.
Several of these protocols achieve lower spacetime overhead than magic state cultivation, partly because the conversion from the QEC code used for the cultivation stage to the final QEC code is more efficient than the conversion from the color code to the grafted code in magic state cultivation.
However, their reliance on non-local connectivity limits their applicability to hardware architectures that do not support non-local operations, such as nearest-neighbor superconducting architectures.
In this work, we address the same bottleneck under the square-grid connectivity constraint: MSC-LS efficiently transfers the magic state from the color code to the rotated surface code using lattice surgery and code expansion.

%% file: 4-proposal.tex
\section{Magic state cultivation with lattice surgery}
\label{sec:msc-ls}

\begin{figure}[t!]
\centering
\includegraphics[width=8.6cm]{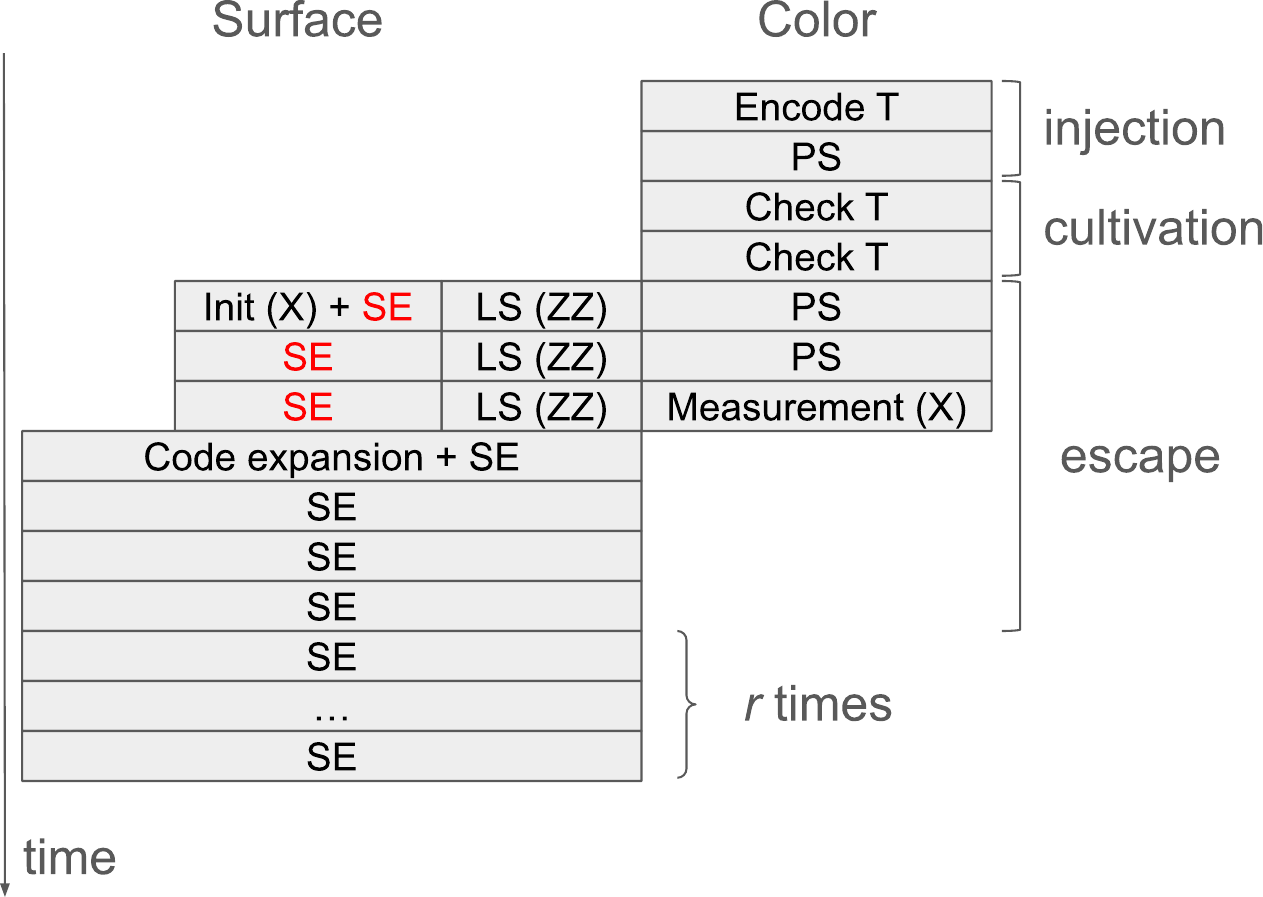}
\caption{
Magic state cultivation with lattice surgery (MSC-LS) for $d_{\mathrm{color}} = 3$.
The left area represents the spatial resource (physical qubits) used by the rotated surface code, and the right area represents those used by the color code.
``Encode T'' denotes the non-fault-tolerant encoding of a magic state into the distance-3 2D color code.
Two consecutive ``Check T'' denote the application of two Hadamard tests.
``LS (ZZ)'' denotes the $ZZ$ measurement implemented by lattice surgery.
``PS'' denotes postselection, and ``SE'' denotes syndrome extraction.
The detectors obtained from the syndrome extractions marked in red are used for early rejection via the lookup table.}
\label{fig:msc-ls-protocol}
\Description{Timeline-style schematic of magic state cultivation with lattice surgery for d_color equal to 3.
The figure is split into two vertical regions labeled Surface on the left and Color on the right, representing the spatial resources used by the rotated surface code and the color code.
Time progresses downward.
The upper part contains the injection and cultivation stages, with an Encode T block followed by a postselection block and two Check T blocks on the color code side, and corresponding lattice surgery and postselection blocks on the surface code side.
The lower part shows the escape stage, beginning with code expansion and followed by three syndrome extraction blocks, with additional syndrome extraction rounds repeated r times.
Syndrome extraction blocks before code expansion are highlighted in red to indicate the rounds whose detectors are used for early rejection.}
\end{figure}

In this section, we describe our proposed protocol: \textit{magic state cultivation with lattice surgery} (MSC-LS).
\autoref{fig:msc-ls-protocol} depicts the protocol with $d_{\mathrm{color}} = 3$.
The protocol shares the injection and cultivation stages with the original magic state cultivation (MSC).
Afterward, the magic state is transferred to the rotated surface code of distance $d_{\mathrm{intermediate}}$ using lattice surgery~\cite{Horsman2012, poulsen2017fault}.
This transfer requires $d_{\mathrm{color}}$ rounds of syndrome extraction.
At the end of the lattice surgery procedure, we perform additional postselection using a lookup table that stores error syndromes that are available at the stage.
We then perform code expansion, followed by $d_{\mathrm{color}}$ rounds of syndrome extraction to stabilize the code.
Finally, we compute the complementary gap and perform postselection based on its value.
We perform $r$ rounds of syndrome extraction while waiting for the QEC decoder to compute the complementary gap.
Similar to the magic state cultivation case, the fault distance of the resulting magic state depends on the complementary gap threshold.

Our protocol differs from MSC in three key aspects: (1) adopting the space-efficient rotated surface code as the final QEC code, (2) using an intermediate code distance during the lattice surgery procedure to reduce the spatial overhead in the early stages, and (3) incorporating the lookup table to enable early rejection.
Although $d_\mathrm{color}$ additional rounds of syndrome extraction are needed for code stabilization after code expansion, the spatial overhead reduction and early rejection outweigh the additional time overhead, as shown later in \autoref{sec:performance-evaluation}.

In Ref.~\cite{Gidney2024MagicStateCultivation}, Gidney \textit{et al.} considered the use of lattice surgery in the escape stage but observed that, in their setting, lattice surgery preserved the fault distance without allowing the code distance to increase during the lattice surgery procedure.
Consequently, the code distance remained $d_\mathrm{color}$ rather than increasing to the larger target distance required for subsequent error correction, leading to logical error probabilities that were orders of magnitude higher than the target state error probability.

In our protocol, lattice surgery likewise preserves the fault distance $d_\mathrm{color}$.
We then perform $d_\mathrm{color}$ additional rounds of syndrome extraction following lattice surgery.
Instead of immediately transitioning to standard error correction---which would reduce the effective fault distance to $\lfloor (d_\mathrm{color}+1)/2 \rfloor$---we use the syndrome information from this stabilization period in the complementary gap calculation for final postselection.
Final postselection based on the complementary gap prevents degradation of the effective fault distance during this stabilization period, and as shown later in \autoref{sec:performance-evaluation}, the resulting logical error probabilities remain comparable to those of MSC.

We detail the lattice surgery and code expansion procedures in the following subsections.
We describe how to construct and use the lookup table in \autoref{subsec:msc-ls-lookup-table}.

\subsection{Lattice surgery}
\label{subsec:msc-ls-ls}
In the lattice surgery procedure, we teleport the magic state from the 2D color code to the rotated surface code of distance $d_{\mathrm{intermediate}}$, following the spirit of the zero-level distillation protocol~\cite{itogawa2025efficient}.
Specifically, we encode $\ket{+}_L$ into the rotated surface code, perform a $Z_LZ_L$ measurement via lattice surgery, and finally measure all physical qubits of the color code in the $X$ basis.

\begin{figure}[!t]
\centering
\includegraphics[width=6.5cm]{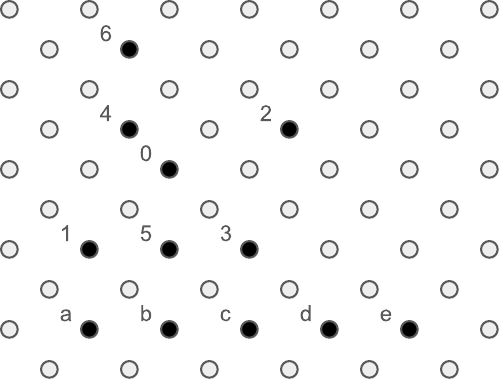}
\caption{
Physical qubit positions.
Qubits $0$--$6$ are data qubits of the distance-3 2D color code, and qubits $a$--$e$ are data qubits on the top edge of the rotated surface code.
}
\label{fig:msc-ls-qubit-coordinates}
\Description{Physical qubit positions.
Qubit 0 is at (3, 13), qubit 1 is at (1, 15), qubit 2 is at (6, 12), qubit 3 is at (5, 15), qubit 4 is at (2, 12), qubit 5 is at (3, 15), and qubit 6 is at (2, 10).
Qubit a is at (1, 17), qubit b is at (3, 17), qubit c is at (5, 17), qubit d is at (7, 17), and qubit e is at (9, 17).
Two qubits can interact when they are at positions (x, y) and (x + 1, y - 1), for any x and y.}
\end{figure}

\begin{figure*}[!t]
\centering
\includegraphics[width=15cm]{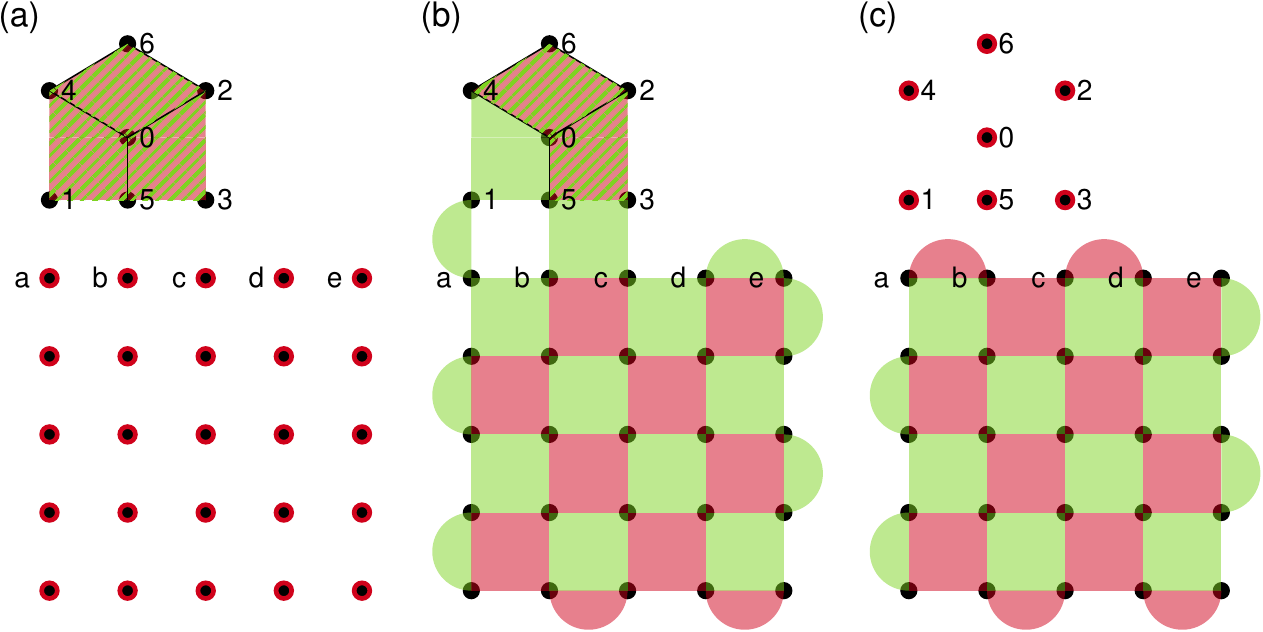}
\caption{
Lattice surgery operation to teleport a magic state from the 2D color code of distance 3 to the rotated surface code of distance 5.
$X$ stabilizers are colored red, and $Z$ stabilizers are colored green.
A red circle on a physical qubit represents initialization or measurement of a physical qubit in the $X$ basis.
(a) The magic state is encoded in the color code, and each physical qubit in the rotated surface code is initialized to $\ket{+}$.
(b) $Z$ stabilizers on the merged boundaries ($Z_{1a}$, $Z_{35bc}$, and $Z_{de}$) are measured three times. Meanwhile, all stabilizers on both codes are measured, except $X_{0145ab}$.
(c) Physical qubits on the color code are measured in the $X$ basis. The $X$ stabilizers on the top boundary of the surface code ($X_{ab}$ and $X_{cd}$) are recovered. $X$ stabilizers $X_{0145ab}$, $X_{0235}$, and $X_{0246}$ are measured, and the outcomes are subject to postselection. $X_{146}$ is the logical $X$ measurement outcome.
}
\label{fig:msc-ls-ls}
\Description{Three panels labeled (a), (b), and (c) showing the progression of a lattice surgery operation between a small color code patch and a rotated surface code patch.
In panel (a), the color code patch appears above the rotated surface code patch. The color code qubits labeled 0 through 6 form a small polygonal patch, and the surface code qubits labeled a through e lie along the top boundary of a rectangular surface code array below.
In panel (b), the color code patch is connected to the top boundary of the surface code patch by the green Z stabilizers Z_{1a}, Z_{35bc}, and Z_{de}.
In panel (c), the color code patch is removed by transversal X measurement, leaving only the labeled qubit positions 0 through 6 above the surface code patch, while the surface code patch remains below.
The figure shows the initial separated layout, the merged layout during lattice surgery, and the final layout after measurement of the color code qubits.}
\end{figure*}

In this subsection, we focus on the case where $d_\mathrm{color} = 3$, and we leave the implementations for larger color code distances to future work.
We denote the physical qubits on the color code as qubits 0--6, and physical qubits on the top edge of the rotated surface code as $a$--$e$ (\autoref{fig:msc-ls-qubit-coordinates}).
\autoref{fig:msc-ls-ls} illustrates the sequence for teleporting a magic state from the color code of distance 3 to the rotated surface code of distance 5.
In the first step, the magic state is encoded in the color code prepared during the injection and cultivation stages discussed in \autoref{subsec:preliminary-magic-state-cultivation}, and each physical qubit in the rotated surface code is initialized to $\ket{+}$ (\autoref{fig:msc-ls-ls} (a)).

In the second step, $Z$ stabilizers on the merged boundaries are measured (\autoref{fig:msc-ls-ls} (b)).
Each measurement is performed three times to detect measurement errors.
Meanwhile, stabilizers on both codes are measured.
Note that, during the construction of the merged code, a four-weight $X$ stabilizer $X_{0145}$ becomes a six-weight $X$ stabilizer $X_{0145ab}$ because $X_{0145}$ and $Z_{1a}$ do not commute.
Since performing a six-weight syndrome extraction is costly, we skip syndrome extraction for $X_{0145ab}$.

Lastly, physical qubits on the color code are measured in the $X$ basis (\autoref{fig:msc-ls-ls} (c)). 
We also recover the $X$ stabilizers along the top edge of the surface code. 
These yield $X$ syndrome measurement outcomes $X_{0145ab}$, $X_{0235}$, and $X_{0246}$, along with the logical $X$ measurement outcome $X_L = X_{146}$.
If the logical $X$ measurement outcome is 1, then we apply the logical $Z$ operator to the magic state encoded in the rotated surface code.
This can be performed classically using the Pauli feedforward mechanism~\cite{riesebos2017pauli}.

In the second step, we use the superdense cycle~\cite{gidney2023new} for syndrome extraction on the color code to measure corresponding $X$ and $Z$ stabilizers simultaneously.
Specifically, in the first syndrome extraction round, we check the following stabilizers: (1) $Z_{0145}$, (2) $Z_{0235}$ and $X_{0235}$, and (3) $Z_{0246}$ and $X_{0246}$.
In the second round, we check (4) $Z_{0145}$, (5) $Z_{0235}$, and (6) $Z_{0246}$.
In the third step of the lattice surgery procedure, we derive (7) $X_{0145ab}$, (8) $X_{0235}$, and (9) $X_{0246}$ from destructive $X$ measurement outcomes on the color code and the syndrome measurement outcome of $X_{ab}$.
In summary, we perform two rounds of syndrome extraction for each stabilizer on the color code except $X_{0145}$, and one round for $X_{0145ab}$ in this procedure.

The lattice surgery procedure performs a $Z_LZ_L$ measurement followed by an $X_L$ measurement.
During the $Z_LZ_L$ measurement, $Z$ errors commute with the logical observable of interest and therefore do not alter the measurement outcome.
Consequently, color code $X$ stabilizer measurements performed prior to the final destructive transversal $X$ measurements are not required to preserve the fault distance.
This explains why skipping the syndrome extraction for $X_{0145ab}$ in the first round does not degrade the fault distance.

Nevertheless, measuring additional $X$ stabilizers can still reduce the logical error probability by detecting intermediate accumulations of $Z$ errors before they form a logical operator.
Whether this reduction materializes in practice depends on the syndrome extraction cost.
Accordingly, we skip $X_{0145ab}$ and retain the other $X$ stabilizer checks in the first round.

During the sequence, color code syndrome extractions are performed in an error-detecting manner to preserve the fault distance of the magic state.
This means that if any error syndrome is nontrivial, the distillation process is rejected.
The same applies to the parities of the $Z$ stabilizers on the merged boundaries.
By contrast, error correction is performed for stabilizers on the rotated surface code.
This is justified when $d_{\mathrm{intermediate}}$ is larger than the fault distance.
As we use a MWPM decoder that requires a detector error model compatible with matching decoders, we provide the decoder with a modified detector error model in which error vertices corresponding to errors on color code qubits are excluded.
These errors are handled purely via error detection and postselection (as described above), rather than correction through matching.

This detector error model modification affects only the decoder input, and we do not modify the underlying circuit-level noise model.
For example, later in this paper (\autoref{sec:performance-evaluation}), we perform numerical simulation using Stim~\cite{gidney2021stim} under uniform depolarizing circuit noise.
There, we maintain two detector error models for the simulation: the complete detector error model and the modified detector error model that is compatible with matching decoders.
We use the former detector error model for sampling, and we give the latter to the PyMatching decoder~\cite{higgott2025sparse}; therefore, all physical error mechanisms are fully represented in the simulation.

The initial measurement outcomes for the $X$ stabilizers introduced in the last step of lattice surgery (except for $X_{ab}$) are non-deterministic.
Hence, extra rounds of syndrome extraction are required to stabilize the code; in \autoref{fig:msc-ls-protocol}, three rounds of syndrome extraction are performed after code expansion.
This is different from the grafted code case, where all the syndrome measurement outcomes are deterministic at the end of the escape stage, and thus no additional rounds are needed (\autoref{fig:preliminary-msc-protocol}).

\subsection{Code expansion}
\label{subsec:msc-ls-code-expansion}
In the code expansion procedure, we increase the distance of the rotated surface code from $d_{\mathrm{intermediate}}$ to $d_{\mathrm{final}}$.
The primary motivation for code expansion is to reduce the qubit count during the early stages of distillation.
Specifically, by teleporting the magic state into a smaller rotated surface code, we reduce the number of physical qubits required during the lattice surgery procedure.
Another advantage of using a small code distance during the lattice surgery procedure is that it shortens the logical $Z$ operator of the rotated surface code, thereby reducing the number of measurements subject to postselection (e.g., $Z_{de}$ in \autoref{fig:msc-ls-ls} (b)) and thus increasing the distillation success probability.

\begin{figure}[t!]
\centering
\includegraphics[width=8cm]{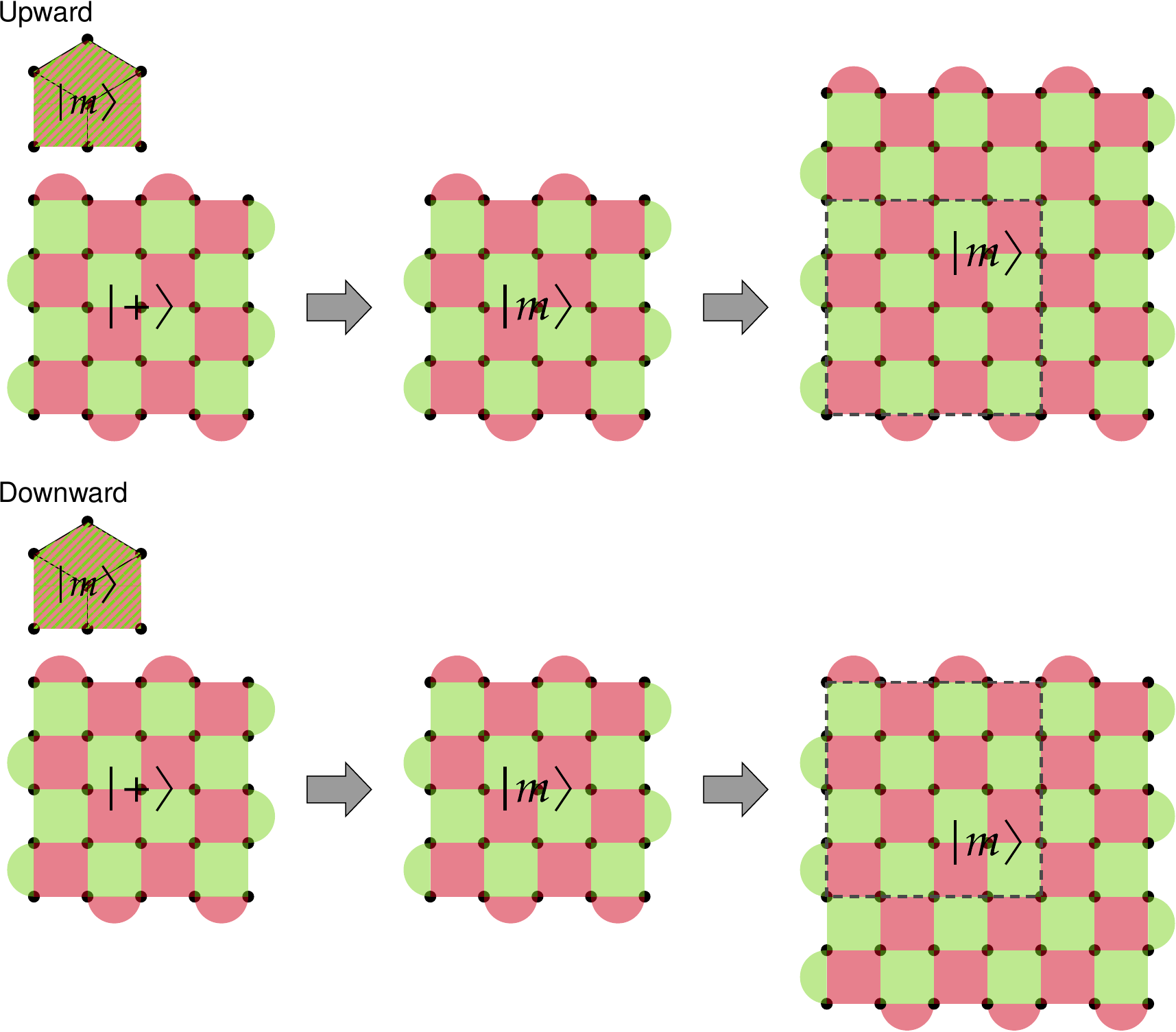}
\caption{
Upward surface code expansion (top) and downward surface code expansion (bottom).
$\ket{m}$ denotes a magic state.
}
\label{fig:msc-ls-upward-downward-code-expansion}
\Description{Two three-step sequences comparing upward and downward surface code expansion.
In both sequences, the color code qubits are initially positioned above a small rotated surface code patch.
The color code is then removed and the surface code patch is expanded.
The top row, labeled Upward, shows the case where the bottom edge of the surface code patch is kept and the patch grows upward.
The bottom row, labeled Downward, shows the case where the top edge of the surface code patch is kept and the patch grows downward.}
\end{figure}

As illustrated in \autoref{fig:msc-ls-upward-downward-code-expansion}, after performing lattice surgery, the surface code patch can be expanded upward or downward.
While these schemes have the same spacetime overhead for a single distillation trial, upward expansion has a packing advantage because the physical qubits used for the color code can be reused for the final surface code patch when $d_{\mathrm{final}}$ is sufficiently large.
Nevertheless, this work focuses on the cost of a single distillation trial; optimizing how multiple trials are packed, scheduled, and repeated to achieve a high overall success probability is a system-level scheduling problem beyond the scope of this work.
For simplicity of implementation, we therefore focus on downward expansion and leave upward expansion and multi-trial scheduling for future work.

\begin{figure}[t!]
\centering
\includegraphics[width=8cm]{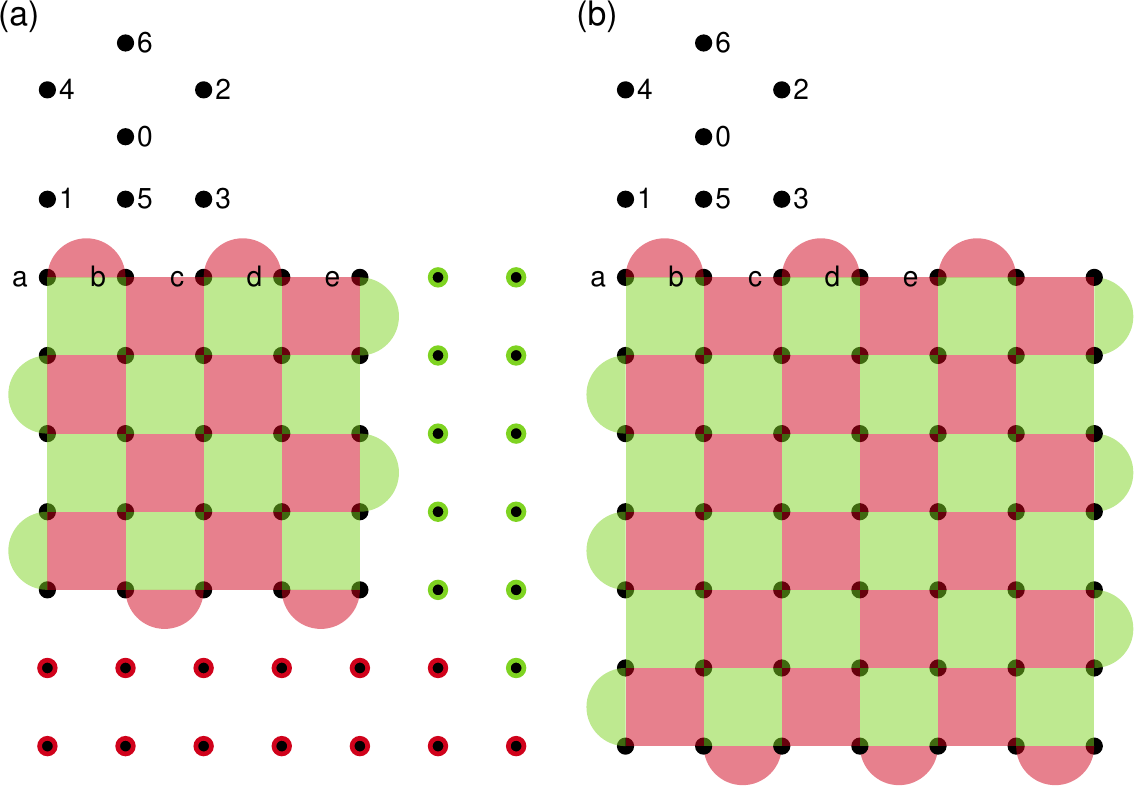}
\caption{
Surface code expansion from distance 5 to distance 7.
A red (green) circle on a physical qubit represents a $\ket{+}$ ($\ket{0}$) initialization.
(a) Additional physical qubits are initialized to $\ket{0}$ or $\ket{+}$ according to their locations.
(b) Additional stabilizers are placed.
}
\label{fig:msc-ls-code-expansion}
\Description{
Two panels labeled (a) and (b) showing surface code expansion from distance 5 to distance 7.
In panel (a), additional qubits are placed around the original patch.
The bottom-left added qubits are initialized in the X basis, and the top-right added qubits are initialized in the Z basis.
In panel (b), additional stabilizers are placed so that the enlarged distance-7 rotated surface code patch is formed.}
\end{figure}

We use Li's protocol~\cite{li2015magic} for the rotated surface code.
\autoref{fig:msc-ls-code-expansion} illustrates the code expansion sequence.
We initialize each physical qubit to $\ket{0}$ or $\ket{+}$, depending on its location, and perform surface code syndrome extraction.

Similar to the discussion in \autoref{subsec:msc-ls-ls}, because the initial syndrome measurement outcomes of some of the newly introduced stabilizers are non-deterministic, several rounds of surface code syndrome extraction are needed to stabilize the code.
This can also be done in parallel with the code stabilization required for lattice surgery.
Hence, we perform $d_{\mathrm{color}}$ rounds of syndrome extraction after code expansion (see \autoref{fig:msc-ls-protocol} for an example).

\subsection{Lookup table}
\label{subsec:msc-ls-lookup-table}
After code expansion, we perform $d_{\mathrm{color}}$ rounds of syndrome extraction to stabilize the code, and $r$ rounds of syndrome extraction while waiting for the decoder to compute the complementary gap.
In this subsection, we describe the lookup table, a data structure used to reject distillation attempts before code expansion.
This reduces the average distillation overhead by allowing us to reject such attempts without waiting for $d_{\mathrm{color}} + r$ rounds.

A lookup table contains syndromes obtained prior to code expansion that would likely be rejected by the final postselection.
At the end of the lattice surgery procedure (before code stabilization), we check whether the obtained error syndrome appears in the table; if so, we reject the distillation attempt.
This is equivalent to predicting whether the logical error probability of the resulting magic state is larger than a certain threshold given the error syndrome.

We expect that whether an error syndrome appears in a lookup table can be checked immediately, because it is significantly simpler than calculating the complementary gap.
Of course, the feasibility of the lookup table depends on the number of entries and the accuracy of the predictions.
We note that our protocol is suitable for the lookup table approach, since $d_{\mathrm{intermediate}}$ is generally small and the error syndromes obtained prior to code expansion are limited in length, resulting in fewer table entries.

\begin{figure}[!t]
  \captionsetup{type=algorithmfigure}
  \hrule \vspace{10pt}
  \caption{Construct a lookup table.}
  \label{alg:construct-lookup-table}
  \hrule \vspace{2pt}
  \begin{algorithmic}
    \State $table = \{\}$
    \State $k \gets $ the size of a syndrome obtained before code expansion
    \For{$shot$ \textbf{in} $shots$}
      \State $synd \gets \text{the error syndrome for } shot$
      \State $pattern \gets synd[0..k]$
      \State $gap \gets$ the complementary gap of $synd$
      \If {$table$ has $pattern$}
        \State $(num\_valid, num\_invalid) \gets table.get(pattern)$
      \Else
        \State $(num\_valid, num\_invalid) \gets (0, 0)$
      \EndIf
      \State $valid \gets$ \textbf{true} if the resulting magic state is correct
      \If {$gap \ge g_\mathrm{th}$ \textbf{and} $valid$}
        \State $num\_valid \pluseq 1$
      \Else
        \State $num\_invalid \pluseq 1$
      \EndIf
      \State $table.set(pattern, (num\_valid, num\_invalid))$
    \EndFor
  \State{Remove entries from $table$ where $num\_valid > 0$.}
  \State{Remove entries from $table$ where $num\_invalid < c_\mathrm{th}$.}
  \State \textbf{return} $table$
  \end{algorithmic}
  \vspace{2pt} \hrule
\end{figure}

\autoref{alg:construct-lookup-table} presents the algorithm for constructing a lookup table.
The algorithm has three inputs: a list of distillation attempts $shots$, the complementary gap threshold $g_\mathrm{th}$, and the threshold on the number of invalid cases $c_\mathrm{th}$.

When the target logical error probability is very small (equivalently, when the target complementary gap is large), it may be that any distillation attempt with a nontrivial syndrome before the code expansion should be rejected.
In this case, we do not store entries in the lookup table.
We instead set a boolean flag $remove\_all\_nontrivial$ in the lookup table.
If the flag is set, the lookup table rejects all nontrivial syndromes.

The lookup table allows us to predict whether the logical error probability of the resulting magic state exceeds a certain threshold, given an error syndrome, while requiring far less computation than calculating the complementary gap.
More sophisticated logic than a simple existence check may further reduce memory requirements and improve prediction accuracy.
In this study, we employ this simple implementation and leave further investigation for future work.
As shown later in \autoref{sec:performance-evaluation}, under our simulation setting with a high complementary gap threshold, a lookup table with 160 thousand entries reduces the average spacetime overhead of distillation by 15\%.
For more memory-constrained systems, a smaller lookup table can be used.

%% file: 5-performance-evaluation.tex
\section{Numerical simulation}
\label{sec:performance-evaluation}

This section presents a performance evaluation of magic state cultivation with lattice surgery (MSC-LS).
We use Stim~\cite{gidney2021stim} and PyMatching~\cite{higgott2025sparse} for the evaluation.
Our implementation is available at \hiddenforreview{\url{https://github.com/yutakahirano/msc-ls/}}.
Because Stim can only simulate Clifford circuits, we replace $T$ gates in the injection and cultivation stages with $S$ gates.
This modification causes the distillation circuit to output $S\ket{+}$ instead of $T\ket{+}$.
Gidney~\textit{et al.}~\cite{Gidney2024MagicStateCultivation} conjecture that cultivating $T\ket{+}$ has a logical error probability approximately twice as high as cultivating $S\ket{+}$.
We note that, in the parameter regime considered below (\autoref{subsec:performance-evaluation-simulation-settings}), namely $d_\mathrm{color} = 3$ and $p_\mathrm{phys} = 10^{-3}$, the validity of this approximation for the injection and cultivation stages is supported by statevector simulations reported in Ref.~\cite{Gidney2024MagicStateCultivation}.
This approximation may also quantitatively affect complementary-gap-based postselection, and thus the final logical error probability and success probability.
Nevertheless, we adopt it in this section because Clifford simulation with this approximation is standard practice in MSC-style analyses.
Recently, several techniques~\cite{wan2025simulate, surti2025efficient, li2025soft} have been proposed to enable simulations of MSC-like protocols without this approximation.
Applying such techniques to MSC-LS is an important direction for future work.

\subsection{Simulation settings} \label{subsec:performance-evaluation-simulation-settings}
We evaluate the performance of MSC-LS using several parameters.
These include the color code distance $d_{\mathrm{color}} = 3$, the intermediate rotated surface code distance $d_\mathrm{intermediate}$, the final rotated surface code distance $d_\mathrm{final}$, and the number of rounds of syndrome extraction $r$ performed while waiting for the complementary gap computation.
Simulations are carried out under uniform depolarizing circuit noise of strength $p_\mathrm{phys}$.
Unless otherwise specified, we set $d_\mathrm{final} = 11$, $r = 10$, and $p_\mathrm{phys} = 10^{-3}$.
For the original magic state cultivation, we use a grafted code distance of $d_\mathrm{grafted} = 15$, following the observation of Gidney~\textit{et al.}~\cite{Gidney2024MagicStateCultivation} that the logical error probability of the distance-15 grafted code is comparable to that of the distance-11 rotated surface code.

We take a ``round'' as the unit of distillation time.
A round is a series of circuit layers that corresponds to one round of syndrome extraction in the 2D color code or the rotated surface code.
This unit allows us to compare the spacetime overhead of distillation with magic state cultivation.
\autoref{alg:performance-evaluation-spacetime-cost} presents the algorithm for calculating the spacetime overhead of distillation.

\begin{figure}[!t]
  \captionsetup{type=algorithmfigure}
  \hrule \vspace{10pt}
  \caption{Calculate the spacetime overhead}
  \label{alg:performance-evaluation-spacetime-cost}
  \hrule \vspace{2pt}
  \begin{algorithmic}
    \State $cost \gets 0$
    \For{each round}
      \State $q \gets$ the number of qubits used in this round
      \State $p \gets$ the success probability of this round
      \State $cost \gets (cost + q) / p$ 
    \EndFor
    \State \textbf{return} $cost$
  \end{algorithmic}
  \vspace{2pt} \hrule
\end{figure}

\subsection{Results}
\label{subsec:performance-evaluation-results}

\begin{figure}[t!]
\centering
\includegraphics[width=8cm]{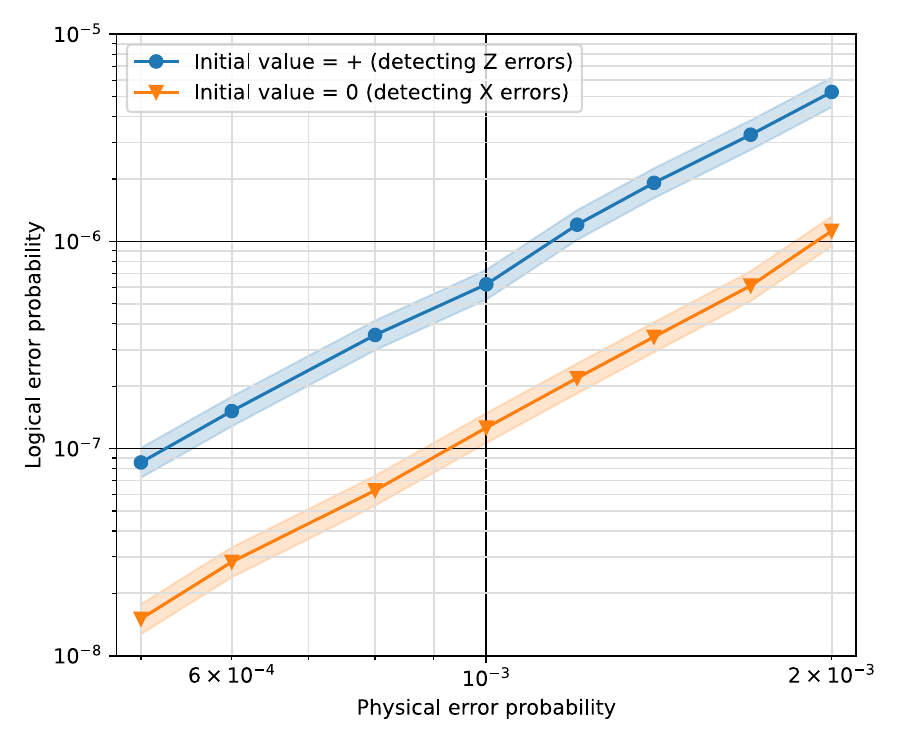}
\caption{Logical $X$ and $Z$ error probabilities introduced by the lattice surgery procedure.}
\label{fig:performance-evaluation-error-detection}
\Description{Fully described in the text.}
\end{figure}

\autoref{fig:performance-evaluation-error-detection} illustrates $X$ and $Z$ logical error probabilities resulting from the lattice surgery procedure described in \autoref{subsec:msc-ls-ls} as a function of $p_\mathrm{phys}$.
Here, to focus on the logical errors caused by lattice surgery, the simulation runs the following sequence.
First, an ideal $\ket{+}_L$ ($\ket{0}_L$) state is encoded in the distance-3 2D color code.
Next, the lattice surgery procedure described in \autoref{subsec:msc-ls-ls} is performed to teleport the quantum state to the rotated surface code of distance $d_\mathrm{intermediate} = 5$.
Finally, we apply one round of noise-free syndrome extraction on the rotated surface code, followed by postselection and a noise-free logical measurement in the $X$ ($Z$) basis, after which we verify the outcome.
This evaluates the $Z$ ($X$) logical error probability of the procedure.
Fitting the data to the form $C\,p_{\mathrm{phys}}^{\alpha}$ gives $\alpha \approx 2.92$ for $Z$ errors and $\alpha \approx 3.30$ for $X$ errors, indicating that the procedure preserves a fault distance of 3, consistent with expectations.

\begin{figure}[t!]
\centering
\includegraphics[width=8.5cm]{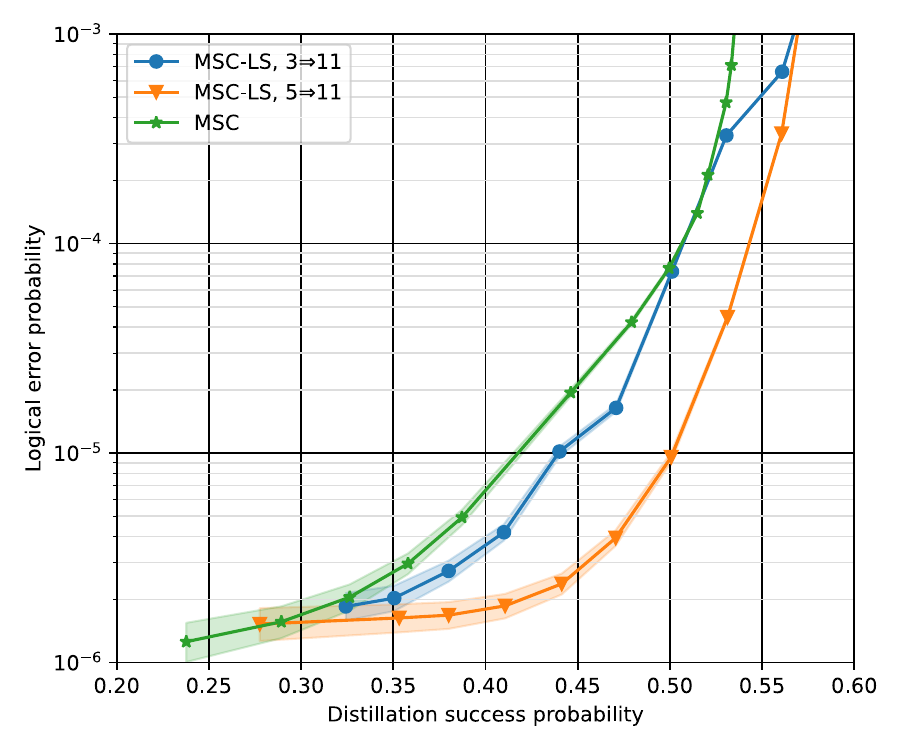}
\caption{
Logical error probability of the resulting magic state as a function of distillation success probability.
Blue: MSC-LS with $d_\mathrm{intermediate} = 3$; orange: MSC-LS with $d_\mathrm{intermediate} = 5$; green: MSC.
Here we do not use the lookup table for MSC-LS.
}
\label{fig:performance-evaluation-end-to-end}
\Description{Plot of logical error probability versus distillation success probability for three protocols.
The horizontal axis shows distillation success probability, and the vertical axis shows logical error probability on a logarithmic scale.
Three increasing curves are shown.
The blue and green curves lie close to each other over much of the plotted range, indicating similar logical error probabilities.
The orange curve lies below them over much of the range, indicating lower logical error probability at the same distillation success probability.
For a fixed logical error probability, the orange curve extends farther to the right than the others, indicating higher distillation success probability.}
\end{figure}

\autoref{fig:performance-evaluation-end-to-end} illustrates the trade-off between the logical error probability and end-to-end distillation success probability of MSC-LS and MSC.
Each point in a series corresponds to a different complementary gap threshold used in the final postselection.
MSC-LS achieves logical error probabilities comparable to those of MSC but outperforms it in end-to-end distillation success probability.
Moreover, MSC-LS with $d_\mathrm{intermediate} = 5$ outperforms MSC-LS with $d_\mathrm{intermediate} = 3$.

The logical error probability in the injection and cultivation stages is $2 \times 10^{-7}$, compared to $8 \times 10^{-7}$ for the lattice surgery procedure (shown in \autoref{fig:performance-evaluation-error-detection}) and $1.6 \times 10^{-6}$ for the end-to-end protocol (shown in \autoref{fig:performance-evaluation-end-to-end}).
This is consistent with the observation in Ref.~\cite{Gidney2024MagicStateCultivation} that the escape stage dominates the logical error budget.

\begin{figure}[t!]
\centering
\includegraphics[width=8.5cm]{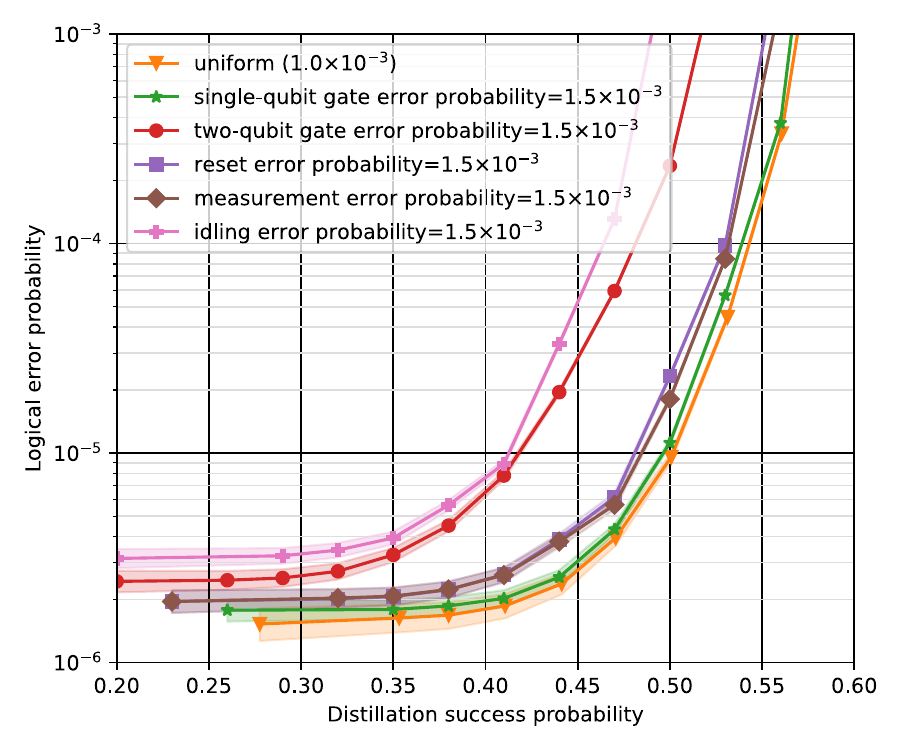}
\caption{
Logical error probability of the resulting magic state as a function of distillation success probability.
The orange curve represents the baseline noise model, which is uniform depolarizing circuit noise of strength $10^{-3}$.
The other curves represent modified noise models in which one error component is set to $1.5 \times 10^{-3}$ while the other error probabilities are kept at their baseline values.
}
\label{fig:performance-evaluation-sensitivity-more}
\Description{Plot of logical error probability versus distillation success probability for the baseline noise model and modified noise models with increased single-qubit gate, two-qubit gate, reset, measurement, and idling error probabilities.
The horizontal axis shows distillation success probability, and the vertical axis shows logical error probability on a logarithmic scale.}
\end{figure}

\begin{figure}[t!]
\centering
\includegraphics[width=8.5cm]{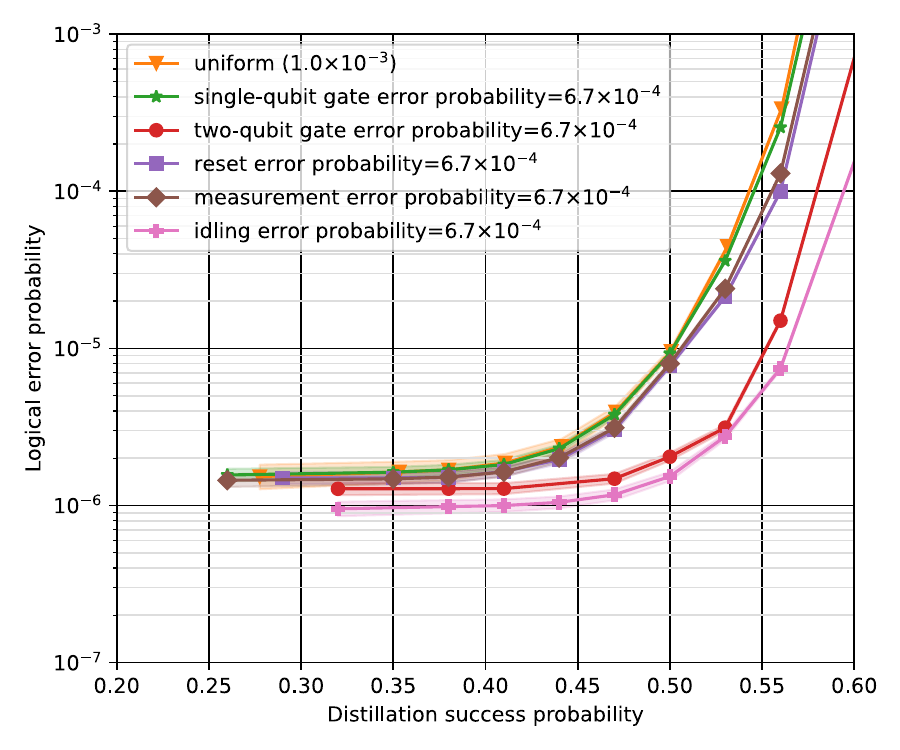}
\caption{
Logical error probability of the resulting magic state as a function of distillation success probability.
The orange curve represents the baseline noise model, which is uniform depolarizing circuit noise of strength $10^{-3}$.
The other curves represent modified noise models in which one error component is set to $6.7 \times 10^{-4}$ while the other error probabilities are kept at their baseline values.
}
\label{fig:performance-evaluation-sensitivity-less}
\Description{Plot of logical error probability versus distillation success probability for the baseline noise model and modified noise models with decreased single-qubit gate, two-qubit gate, reset, measurement, and idling error probabilities.
The horizontal axis shows distillation success probability, and the vertical axis shows logical error probability on a logarithmic scale.}
\end{figure}

To evaluate the sensitivity of these results to the noise model, we vary each error component independently while keeping the others fixed.
\autoref{fig:performance-evaluation-sensitivity-more} and \autoref{fig:performance-evaluation-sensitivity-less} show the resulting logical error probability as a function of the distillation success probability when the error probability of each component is increased and decreased, respectively.
These results show how each error source affects MSC-LS separately, providing a more detailed view of which hardware improvements would most directly improve the performance of the protocol.
In particular, the logical error probability is more sensitive to changes in the idling and two-qubit gate error probabilities than to changes in the other error probabilities.

\begin{figure}[t!]
\centering
\includegraphics[width=8.5cm]{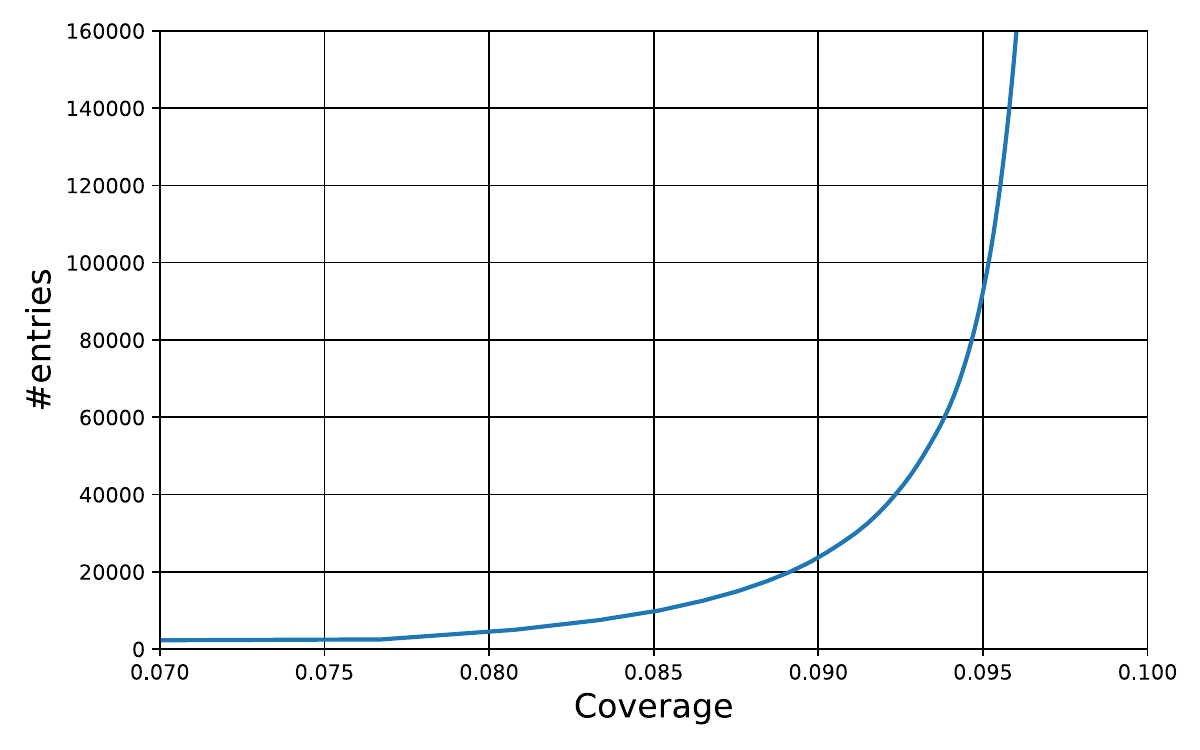}
\caption{
Coverage of the lookup table.
The horizontal axis denotes the ratio of distillation attempts that would be additionally rejected by the lookup table.
The vertical axis represents the number of entries in the table.
}
\label{fig:performance-evaluation-lookup-table-coverage}
\Description{Plot of lookup table size versus coverage.
The horizontal axis shows coverage, and the vertical axis shows the number of entries in the lookup table.
The curve rises gradually at low coverage and then increases sharply near coverage 0.096, reaching about 160,000 entries.}
\end{figure}

\begin{figure}[t!]
\centering
\includegraphics[width=8.5cm]{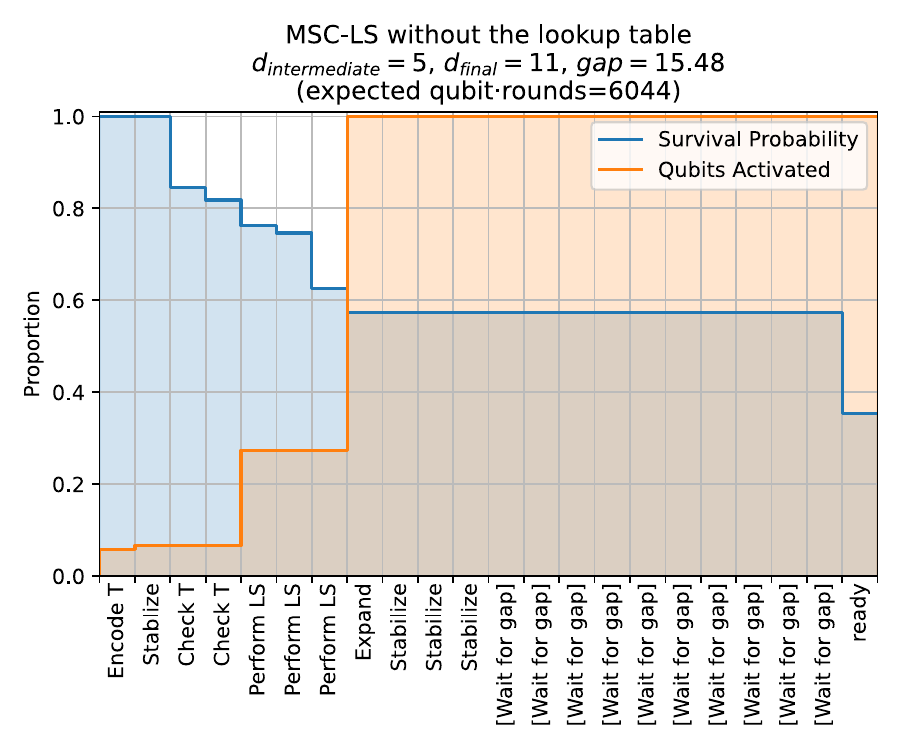}
\caption{
Qubit usage and survival probability during distillation without the lookup table.
Each column represents a round.
The blue region represents the survival probability of distillation.
The orange region represents the physical qubit usage normalized to the maximum qubit usage (241 qubits in this case). 
}
\label{fig:performance-evaluation-distillation-lifetime-without-lookup-table}
\Description{Fully described in the text.}
\end{figure}

\begin{figure}[t!]
\centering
\includegraphics[width=8.5cm]{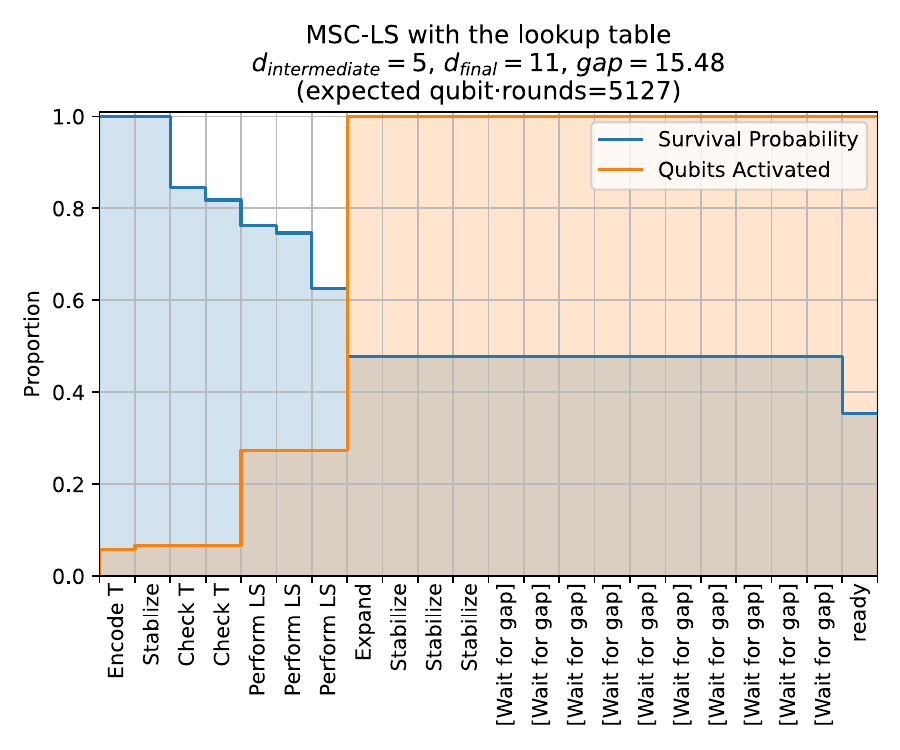}
\caption{
Qubit usage and survival probability during distillation with the lookup table, using the same color scheme as \autoref{fig:performance-evaluation-distillation-lifetime-without-lookup-table}.
}
\label{fig:performance-evaluation-distillation-lifetime-with-lookup-table}
\Description{Combined step plot of qubit usage and survival probability during distillation with the lookup table.
Because additional lookup-table-based postselection is performed before code expansion, the survival probability is lower from that point until the final postselection.
The final survival probability is unchanged.}
\end{figure}

We construct lookup tables using \autoref{alg:construct-lookup-table} with various complementary gap thresholds $g_\mathrm{th}$.
For each $g_\mathrm{th}$ setting, we sample $10^{10}$ shots, with the threshold for invalid cases set to $c_\mathrm{th} = 100$.
\autoref{fig:performance-evaluation-lookup-table-coverage} shows the coverage of the lookup table with $g_\mathrm{th} = 15.48$.
For this table, the offline construction took about 5.7 hours using 90 CPU cores on Intel Xeon Platinum 9242 processors, corresponding to approximately 510 core-hours.
At maximum size, the table contains 159,321 entries and rejects 9.6\% of distillation attempts.
\autoref{fig:performance-evaluation-distillation-lifetime-without-lookup-table} and \autoref{fig:performance-evaluation-distillation-lifetime-with-lookup-table} show the qubit usage and survival probability during distillation.
Without the lookup table (\autoref{fig:performance-evaluation-distillation-lifetime-without-lookup-table}), 43\% of samples are rejected before code expansion, 22\% by the final postselection, and 35\% are accepted.
By contrast, with the lookup table (\autoref{fig:performance-evaluation-distillation-lifetime-with-lookup-table}), 52\% of samples are rejected before code expansion, 13\% are rejected by the final postselection, and 35\% are accepted.
This early rejection reduces the spacetime overhead by 15\% in this setting.


\begin{figure}[t!]
\centering
\includegraphics[width=8.5cm]{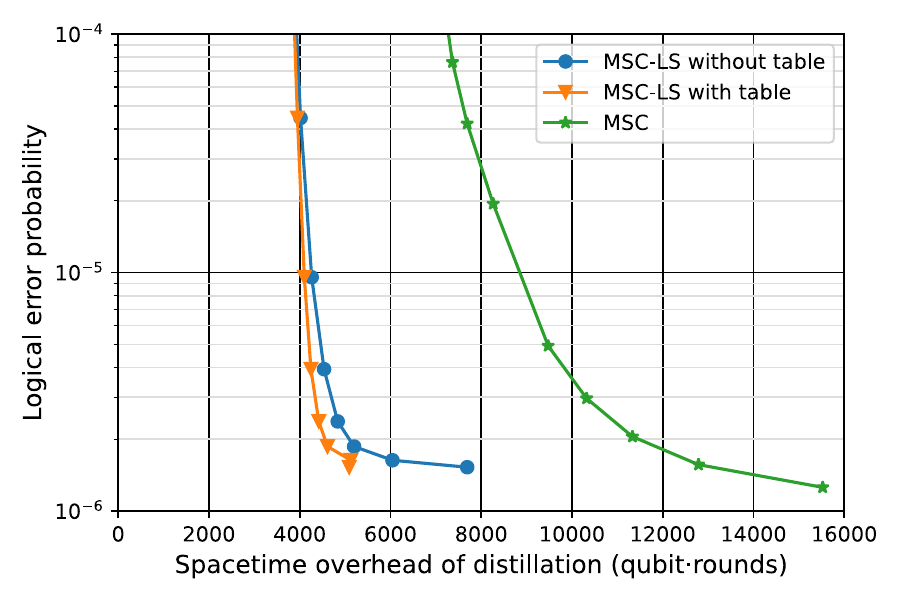}
\caption{
Spacetime overhead of MSC-LS with and without the lookup table and MSC.
}
\label{fig:performance-evaluation-distillation-overhead}
\Description{Plot of logical error probability versus spacetime overhead of distillation for MSC-LS without table, MSC-LS with table, and MSC.
The horizontal axis shows spacetime overhead in qubit-rounds, and the vertical axis shows logical error probability on a logarithmic scale.
All three curves decrease as spacetime overhead increases.
The two MSC-LS curves lie to the left of the MSC curve, and the lookup-table version lies slightly to the left of the version without the table in the lower logical-error region.}
\end{figure}

\autoref{fig:performance-evaluation-distillation-overhead} depicts the spacetime overhead of distillation of MSC-LS and MSC.
It shows that in the logical error probability range of $[10^{-6}, 10^{-5}]$, MSC-LS requires less than half the spacetime overhead required for MSC.
For example, MSC-LS reduces the spacetime overhead by 59\% (55\%) when distilling an $S\ket{+}$ state with a logical error probability of $2 \times 10^{-6}$ ($4 \times 10^{-6}$).
The lookup table performs more effectively in the lower range of logical error probability, which is expected, given that postselection is performed more strongly in this range.

\autoref{fig:performance-evaluation-lookup-table-coverage} also shows that the lookup table is useful even when the number of entries in the table is limited.
For example, with only 5,000 entries, the coverage reaches 8\%.
Hence, the lookup table is useful even in memory-constrained systems.
The number of detectors contained in a table entry is 54 with our simulation setting, and hence each table entry requires only 54 bits and is very lightweight.

%% file: 6-conclusion.tex
\section{Conclusion}
\label{sec:conclusion}

In this study, we proposed magic state cultivation with lattice surgery, an efficient magic state distillation protocol.
The efficiency of our protocol arises from (1) adopting the space-efficient rotated surface code as the final QEC code, (2) using an intermediate code distance smaller than $d_\mathrm{final}$ during the lattice surgery procedure to reduce the spatial overhead in early stages, and (3) incorporating the lookup table to enable early rejection.
Our protocol also provides simplicity, as it avoids the complex grafting procedure and the grafted code.
Numerical simulations show that our protocol achieves a logical error probability for the resulting magic state comparable to that of magic state cultivation, while requiring only about half the spacetime overhead of magic state cultivation.
Although we leave implementations with larger color code distances for future work, the $d_\mathrm{color} = 3$ implementation studied here is already useful in practical regimes, as discussed below.
Therefore, we conclude that magic state cultivation with lattice surgery offers a practical path to more efficient magic state distillation in fault-tolerant quantum computation, thereby reducing the total spacetime cost of $T$-intensive quantum computation.

In this study, we have focused on implementations with $d_\mathrm{color} = 3$.
We see no fundamental obstacles to implementations at larger color code distances.
Whether such implementations are more efficient than magic state cultivation with the corresponding color code distance remains open and is an important direction for future work.
Nevertheless, MSC-LS with $d_\mathrm{color} = 3$ is already relevant in practical regimes.
It prepares a $T$ magic state with a logical error probability of approximately $3 \times 10^{-6}$, which is useful in the megaquop regime~\cite{preskill2025beyond}.
This regime is an important next milestone beyond NISQ and includes many promising applications of early fault-tolerant quantum computation.
Moreover, $T$ magic states prepared by our protocol can be used as inputs to further distillation~\cite{Litinski2019magicstate}.
For example, the $8T$-to-$\mathrm{CCZ}$ distillation protocol~\cite{Campbell2017Unified} can prepare a higher-fidelity $\mathrm{CCZ}$ magic state, which can be used to implement the Toffoli gate.
Since the $8T$-to-$\mathrm{CCZ}$ protocol's leading-order output logical error probability is $28p^2$, $T$ magic states prepared by MSC-LS with $d_\mathrm{color} = 3$ can be used as inputs to prepare $\mathrm{CCZ}$ magic states with a logical error probability on the order of $10^{-10}$.

Another important direction for future work is to develop a more concrete distillation pipeline.
Both MSC and MSC-LS have relatively low end-to-end success probabilities, and the average spacetime cost of distillation is evaluated under the assumption that distillation can be parallelized and pipelined.
For instance, satisfying this assumption requires additional code expansion schemes beyond the downward expansion discussed in \autoref{subsec:msc-ls-code-expansion}.
This capability is critical for using the protocol in compilation schemes that require a small, on-demand distillation factory~\cite{hirano2025locality}.